\documentclass[twocolumn,showpacs,eqsecnum,amsmath,amssymb,prb]{revtex4}

\usepackage{graphicx}
\usepackage{dcolumn}
\usepackage{bm}

\begin{document}


\title { 
The structural phase transition and loss of magnetic moments
in NpO$_2$: \\
{\it ab initio} approach to the crystal and mean field
}

\author {A.V. Nikolaev} 
 \altaffiliation[Also at:]{
Institute of Physical Chemistry of RAS, 
Leninskii prospect 31, 117915, Moscow, Russia 
} 

\author{K.H. Michel} 

\affiliation{%
Department of Physics, University of Antwerp, UIA, 2610, Antwerpen, Belgium
}%

\date{\today}

\begin{abstract} 
We discuss the triple-$\vec{q}^X$ structures 
for the ordered cubic phase of NpO$_2$, which are $Pn{\bar 3}m$ 
and $Pa{\bar 3}$. 
A special care should be
taken to discriminate between these two cases.
We analyze the relevant structure-factor amplitudes
and the effect of domains on resonant X-ray
scattering experiments.
We formulate the technique of multipole expansion of the Coulomb
interaction and use it to study the crystal electric field and
mean field for a number of neptunium many electron configurations 
($5f^3$, $7s5f^3$, $7p5f^3$ and $6d5f^3$) on {\it ab initio} level. 
We have found that the crystal field is rather small (50-150~K).
The direct quadrupole-quadrupole interaction between neighboring
neptunium sites is weak and can not
drive the structural phase transition at 25~K.
We have introduced an effective (enhanced) quadrupole interaction
and considered the interplay between it and the crystal electric
field.
The influence of both interactions on the transition temperature
has been investigated in detail for the $5f^3$ configuration.
We discuss the importance of the intrasite multipole interaction
between three localized electrons ($5f^3$) and a valence electron
on a neptunium site. We show that this interaction 
combined with the symmetry lowering at 25~K may be responsible 
for the loss of the magnetic moments in the ordered phase of NpO$_2$.
\end{abstract} 
 
\pacs{64.70.Kb, 71.70.-d, 71.70.Ch, 75.10.Dg} 

\maketitle

\section {Introduction} 
\label{sec:int} 

Since the discovery of the low temperature phase transition
in NpO$_2$ at $T_c=25.5$~K almost fifty years ago, \cite{Wes} the 
magnetic properties of this compound represent a challenge
for theoretical interpretation. \cite{Erd,San1} 
NpO$_2$ and its neighbor in the actinide series, UO$_2$, 
\cite{Fra,Cac} are both crystallized 
in the cubic fluorite
(CaF$_2$) structure in the high temperature phase.
The temperature dependence of the magnetic susceptibilities of
these compounds is suggestive for a N\'{e}el transition
to an antiferromagnetic state. \cite{Erd2,Fra}
However, while it was found that UO$_2$ orders antiferromagnetically 
($T_N=30.8$~K) with 1.74$\mu_B$/U atom, \cite{Fra} 
no magnetic ordering \cite{Cac2} was
confirmed for NpO$_2$ (see for a review Ref.\ \onlinecite{San1}).
As follows from M\"{o}ssbauer experiments \cite{Fri}
the upper limit on the magnetic moments of Np in the ordered phase
is only 0.01$\mu_B$/Np.
This implies almost a complete quenching of the 3.0$\mu_B$ 
magnetic moment
of Np, which it exhibits in the paramagnetic high temperature phase.
The demagnetization of Np magnetic moments seems to be incompatible
with the widely accepted viewpoint that the Np cations are in the
tetravalent valence state, Np$^{4+}$, 
with three localized 5f electrons
left at each neptunium site. 
With the $5f^3$ configuration, Np becomes
a Kramers ion having a magnetic doublet ($\Gamma_6$, $\Gamma_7$) or 
quadruplet ($\Gamma_8$) ground state. Because of this, a divergent
susceptibility at $T \rightarrow 0$ will be present in any model of
the phase transition. \cite{San1,San2}

To describe the disappearance of magnetic moments of Np
at $T<T_c$
Santini and Amoretti put forward an idea of a magnetic octupole
order parameter, \cite{San2} which on one hand,
 is not invariant under time reversal symmetry 
and, on the other hand, is different from the magnetic dipolar
order parameter which brings about the ordinary magnetic ordering.
The idea of breaking down of Kramers' degeneracy by the magnetic
octupole 
seems to be the only solution  in the framework of the 
$5f^3$ model. \cite{San2,San1}

An alternative explanation was offered by Friedt {\it et al}. \cite{Fri}.
They suggested that 
the magnetic anomaly in NpO$_2$ could be explained by
a splitting of the ground state $\Gamma_8$ quadruplet of cubic symmetry
into two doublets in the crystal field of lower symmetry of the
ordered phase. 
The splitting was ascribed to an internal
distortion of the oxygen sublattice at $T_c$.
This viewpoint was inspired by the experimental observation that 
the phase transition in UO$_2$ is accompanied by an internal 
distortion of the oxygen cube that surrounds the U cation, \cite{Fab}
while the external cubic structure of uranium dioxides survives
the transition. 
However, the mechanism mainly weakens the magnetic response
from the Np sites at 20~K, while the $T \rightarrow 0$
divergence of magnetic susceptibility still persists 
in the model in contradistinction 
with experiment. \cite{Erd2,Cac2} 
On the other hand, no evidence for an internal or external 
crystallographic distortion in NpO$_2$
has been found by synchrotron
experiments. \cite{Man}
Thus, the phase transition in NpO$_2$ appears to
be isostructural like the 
 $\gamma-\alpha$ phase transition in elemental
cerium \cite{Kos,Mal} ($T_c \sim 100$~K)
or the isostructural expansion in YbInCu$_4$ at $T=42$~K. \cite{Sar}
It is worth mentioning that {\it in all these compounds the phase
transition is accompanied by a loss of magnetic moments
in the ordered phase.}

However, very recent resonant X-ray scattering (RXS) experiments
at the Np $M_{IV}$ and $M_{V}$ edges in NpO$_2$ 
indicated an unexpected
result: the phase transition is not isostructural. \cite{Pai} 
In the
low temperature phase a long range order of Np electric quadrupoles
was revealed by the growth of superlattice Bragg peaks. \cite{Man,Pai}
The space symmetry of the ordered phase was identified 
as $Pn{\bar 3}m$. \cite{Pai}
The symmetry lowering is a special one. The centers of mass positions
of neptunium and oxygen remain in the cubic CaF$_2$ structure
as in the high temperature phase. However, electronic quadrupoles
of the Np sites have four different orientations, which allows us
to distinguish four different sublattices of Np cations. \cite{Pai} 

This experimental finding gives rise to a question of correlations
between structural and magnetic properties in solids, which we
have investigated theoretically in our model of
the $\gamma-\alpha$ phase transition in Ce. \cite{NM1,NM2,NM3} 
Unlike NpO$_2$, pristine cerium is a metal and the phase transition
there at normal pressure is accompanied by a huge volume 
change. \cite{Kos,Mal}
The volume anomaly in NpO$_2$ (contraction) is only 0.018\%. \cite{Man}
However, the disappearance of the magnetic moments and the ``isostructural"
character of the phase transition makes them similar.
In this respect it is interesting to notice that in our model for Ce
we have predicted the $Pa{\bar 3}$ space symmetry for the ordered 
$\alpha$ phase \cite{NM1} which is very close to
the $Pn{\bar 3}m$ structure reported for NpO$_2$ 
in Ref.\ \onlinecite{Pai}.
The active irreducible representation also belongs to the $X$
point of the Brillouin zone (BZ).
While the existence of superstructure reflections for 
$\alpha$-Ce remains
an open question which has to be investigated 
experimentally, \cite{NM1,NM2,NM3}
here we want to apply our theoretical concepts for the study 
of the crystal- and mean-field in neptunium dioxide.

We will use the technique of multipole expansion of the Coulomb
interaction. \cite{NM1,NM2,NM3,NM4}
The multipole expansion represents a unified description of the
crystal field effects and atomic term splitting.
The intrasite Coulomb repulsion which is responsible for Hund's rules,
the spin-orbit coupling and the crystal field effects 
are included on equal footing.
(For a single site the technique is equivalent to the classical description
of atomic terms. \cite{CS})
In order to calculate the corresponding electronic spectra we use
many Slater determinants, which indicates that our scheme is
a genuine many electron approach \cite{NM3,NM4} corresponding to the
configuration interaction (CI).
(Notice that ordinarily used single-determinant Hartree-Fock approach is 
not sufficient for these purposes).

The paper is organized as follows.
In Sec.~II we discuss the triple-$\vec{q}^X$ structures 
($Pn{\bar 3}m$ or $Pa{\bar 3}$) of the ordered phases and
derive the selection rules for resonant X-ray scattering
(RXS) experiments.
In Sec.~III we describe our method for treating
local electron configurations at neptunium sites.
Next (Sec.~IV) we examine thoroughly the case of three localized
$5f$ electrons. In Sec.~V we generalize the $5f^3$ model
by including the multipole interactions with a valence electron
which is present instantaneously on a Np site.
We show that in the framework of this extended model 
the disappearance of the
magnetic moments can be understood and ascribed to the 
trigonal symmetry lowering at 25~K. 
Our conclusions are summarized in Sec.~VI.

\section {Structural phase transitions and RXS experiments}

In this section we discuss the allowed space groups of the
low temperature phase of NpO$_2$. 
Our considerations here are based on a general group
theoretical approach which is independent of the model
assumptions.

In the disordered phase ($T>25$~K) the crystal structure is
cubic ($Fm{\bar 3}m$) and the site group is $O_h$. The
electron density of neptunium
is given by the spherically symmetric component ($Y_0^0$) and the cubic
harmonics $K_{l}(\Omega)$ with $l=4,6$. 
There is no contribution to the density
from the quadrupole functions $Y_{l=2}^{\tau}$,
where $\tau=0$, $(m,c)$ or $(m,s)$, $m=1,2$.
(Here and below we work with the real spherical harmonics
with the phase definition of Ref.~\onlinecite{Bra}.)
At $T_c=25$~K the transition to a new  
phase sets in. The new phase was characterized as a triple-$\vec{q}$
antiferro- quadrupolar ordering of $T_{2g}$ ($\Gamma_{5g}$) 
electric quadrupoles at the Np sites. \cite{Pai}

In real space the ordering is characterized by four different
sublattices of the simple cubic structure.
We label these sublattices which contain the sites (0,0,0)
$(a/2)$(0,1,1), $(a/2)$(1,0,1) and $(a/2)$(1,1,0) by $\{ \vec{n}_p \}$,
$p=1-4$, respectively.
The most significant feature of the ordered phase is the
existence of only one three-fold axis of symmetry $C_3$ at each Np site
which is also a cube diagonal.  
The only quadrupole function compatible with the symmetry
lowering is $Y_2^0(\Omega')$ in the coordinate system where the 
$z'$-axis
coincides with one of the three-fold axes (cube diagonals): 
$[111]$, $[-1,-1,1]$, $[1,-1,-1]$, and $[-1,1,-1]$. 
Consequently, there are four such
functions which are given by
\begin{subequations}
\begin{eqnarray}
 & & {\cal S}_{a}(\Omega)=
 \frac{1}{\sqrt{3}}(Y_2^{1s}(\Omega)+Y_2^{1c}(\Omega)+Y_2^{2s}(\Omega)) ,   
  \quad  \quad  \quad \label{3c.1a}  \\ 
 & & {\cal S}_{b}(\Omega)=
 \frac{1}{\sqrt{3}}(-Y_2^{1s}(\Omega)-Y_2^{1c}(\Omega)+Y_2^{2s}(\Omega)) ,   
  \quad  \quad   \quad \label{3c.1b}  \\ 
 & & {\cal S}_{c}(\Omega)=
 \frac{1}{\sqrt{3}}(Y_2^{1s}(\Omega)-Y_2^{1c}(\Omega)-Y_2^{2s}(\Omega)) ,   
  \quad  \quad   \quad \label{3c.1c}  \\ 
 & & {\cal S}_{d}(\Omega)=
 \frac{1}{\sqrt{3}}(-Y_2^{1s}(\Omega)+Y_2^{1c}(\Omega)-Y_2^{2s}(\Omega)) .   
  \quad  \quad   \quad \label{3c.1d}   
\end{eqnarray}
\end{subequations}
The spherical harmonics $Y_2^{1s}$, $Y_2^{1c}$, $Y_2^{2s}$
belong to a three-dimensional irreducible representation
$T_{2g}$ of $O_h$. They are proportional to 
the Cartesian components $yz$, $zx$, and $xy$.

\begin{figure} 
\vspace{-8mm}
\resizebox{0.48\textwidth}{!}
{ 
 \includegraphics{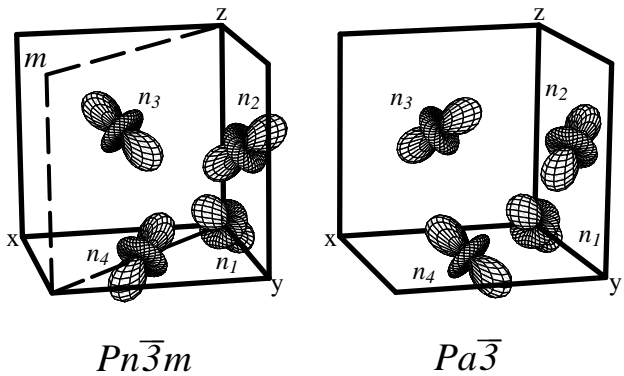} 
} 
\caption{
$Pn{\bar 3}m$ and $Pa{\bar 3}$ structures for quadrupoles of NpO$_2$.
Both structures have the common $S_6$ (or ${\bar 3}$) local site symmetry.
They differ in the way the threefold axes and quadrupoles
are distributed over 4 sublattices $\{ n_p \}$, $p=1-4$,
(see text for details). The $Pn{\bar 3}m$ structure possesses three
mirror planes one of which ($m$) is shown explicitly.
} 
\label{fig_struc} 
\end{figure}

In Ref.~\onlinecite{Pai} the space group of the ordered phase
was identified as $Pn{\bar 3}m$ (No.~224, Ref.~\onlinecite{Tables}).
The corresponding ordering
of ${\cal S}_a-{\cal S}_d$ functions is shown in Fig.~\ref{fig_struc},
left panel.
%
%
However, a close examination reveals that
there is another possibility which was not considered
by the authors of Ref.~\onlinecite{Pai}. This is the $Pa{\bar 3}$
structure (No.~205 of Ref.~\onlinecite{Tables})
depicted in the right panel of Fig.~\ref{fig_struc}.
The $Pa{\bar 3}$ structure also realizes a triple-$\vec{q}$ 
quadrupolar order, the site symmetry of the Np sites is 
$S_6=C_3 \times i$.

The problem of finding the appropriate space group 
can be simplified to the task of distributing 
four orientational functions ${\cal S}_a-{\cal S}_d$ among
four distinct sublattices $\{ \vec{n}_p \}$, $p=1-4$.
Keeping the three-fold rotation symmetry about the $[111]$ axis
and ${\cal S}_a$ for the first sublattice,
we distinguish only three different choices here.
The first is when $(i)$ 
${\cal S}_b$ corresponds to $\{ \vec{n}_4 \}$,
${\cal S}_c-\{ \vec{n}_2 \}$, and ${\cal S}_d-\{ \vec{n}_3 \}$.
This gives the $Pn{\bar 3}m$ structure, Fig.~\ref{fig_struc},
left panel.
Another assignment is $(ii)$
${\cal S}_b-\{ \vec{n}_2 \}$,
${\cal S}_c-\{ \vec{n}_3 \}$, and ${\cal S}_d-\{ \vec{n}_4 \}$.
This scheme corresponds to the $Pa{\bar 3}$ structure as
shown in Fig.~\ref{fig_struc}, right panel.
The third choice is $(iii)$ 
${\cal S}_b-\{ \vec{n}_3 \}$,
${\cal S}_c-\{ \vec{n}_4 \}$, and ${\cal S}_d-\{ \vec{n}_2 \}$.
This is simply another variant (or domain) of the 
$Pa{\bar 3}$ structure.

From the mathematical point of view we are analyzing the
symmetry lowering and the condensation schemes at the $X$ point
of the BZ of the fcc lattice, 
which involves the density components of the $T_{2g}$ symmetry. 
In both cases ($Pn{\bar 3}m$ and $Pa{\bar 3}$)
the triple-$\vec{q}$ mode which drives the structural phase transition
belongs to the $X$ point of BZ $^*\! \vec{q}^X$, 
and involves the three arms of the latter,
$\vec{q}^X_x=(2\pi/a)(1,0,0)$, $\vec{q}^X_y=(2\pi/a)(0,1,0)$,
and $\vec{q}^X_z=(2\pi/a)(0,0,1)$.
To describe the three quadrupolar components 
$S_1=Y_2^{1s}$, $S_2=Y_2^{1c}$, $S_3=Y_2^{2s}$
at site $\vec{n}$ we introduce the functions $S_i(\vec{n})$ ($i=1-3$)
and consider their Fourier transforms,
\begin{eqnarray}
  S_i(\vec{q})=\frac{1}{\sqrt{N}} \sum_{\vec{n}} 
  e^{i\vec{q}\cdot \vec{X}(\vec{n})} S_i(\vec{n}) ,
\quad \label{3c.0}
\end{eqnarray}
where the position vector $\vec{X}$ runs over $N$ sites of 
the face centered cubic (fcc) neptunium lattice.
The little group of $^*\! \vec{q}^X$ is $D_{4h}$ ($4/mmm$). 
At $\vec{q}^X_x$ the functions $S_2$ and $S_3$ belong
to the $E_g$ small representation, while the function
$S_1$ belong to the $B_{2g}$ small representation.
Consequently, we distinguish two different irreducible
representations of the space group $Fm{\bar 3}m$.
The first one, $X_5^+$ in notations of Stokes and Hatch, \cite{Sto} 
comprises
six orientational functions, with two functions from every
arm of $^*\! \vec{q}^X$. The functions are $S_2(\vec{q}^X_x)$,
$S_3(\vec{q}^X_x)$; $S_1(\vec{q}^X_y)$, $S_3(\vec{q}^X_y)$;
$S_1(\vec{q}^X_z)$, and $S_2(\vec{q}^X_z)$.
The second irreducible representation, $X_4^+$, 
has only three components, with
one function from each arm of $^*\! \vec{q}^X$, namely
$S_1(\vec{q}^X_x)$, $S_2(\vec{q}^X_y)$ and
$S_3(\vec{q}^X_z)$.
For the symmetry lowering to the $Pn{\bar 3}m$ structure
we obtain the following condensation scheme:
\begin{eqnarray}
 & & Fm{\bar 3}m: X_4^+  \nonumber \\
 & & ( \langle S_1(\vec{q}^X_x) \rangle=\langle S_2(\vec{q}^X_y) \rangle
 =\langle S_3(\vec{q}^X_z) \rangle=\rho_1 \sqrt{N} \neq 0 ) \nonumber \\
 & & \rightarrow Pn{\bar 3}m  \quad (Z=4) . \label{Pn3m}
\end{eqnarray}
Here $\langle ... \rangle$ stand for a quantum and thermal average
and $\rho_1$ is the order parameter amplitude.
The $X_4^+$ active representation corresponds to the irreducible
star $\{ \vec{k}_{10} \}$ and the loaded one-dimensional 
representation $\hat{\tau}^7$ in Kovalev's notation. \cite{Kov}
There are four domains for the $Pn{\bar 3}m$ structure.

The condensation scheme for the $Pa{\bar 3}$ structure is given by
\begin{eqnarray}
 & & Fm{\bar 3}m: X_5^+ \nonumber \\
 & & ( \langle S_3(\vec{q}^X_x) \rangle=\langle S_1(\vec{q}^X_y) \rangle
 =\langle S_2(\vec{q}^X_z) \rangle=\rho_2 \sqrt{N} \neq 0 ,
 \nonumber \\
 & & \langle S_2(\vec{q}^X_x) \rangle=\langle S_3(\vec{q}^X_y) \rangle
 =\langle S_1(\vec{q}^X_z) \rangle = 0 ) \nonumber \\
 & & \rightarrow Pa{\bar 3} \quad (Z=4) 
 \quad \mbox{domain I} . \label{Pa3}
\end{eqnarray}
Here again $\rho_2$ is the order parameter amplitude.
This condensation scheme corresponds to a domain 
shown in Fig.~\ref{fig_struc}, and there are
eight possible domains of $Pa{\bar 3}$.
In particular, there is a domain $(iii)$ which we have already considered
before (${\cal S}_a-\{ \vec{n}_1 \}$, 
${\cal S}_b-\{ \vec{n}_3 \}$,
${\cal S}_c-\{ \vec{n}_4 \}$, and ${\cal S}_d-\{ \vec{n}_2 \}$)
when we analyzed the variants of
distribution of ${\cal S}_a-{\cal S}_d$ over four sublattices.
This domain is obtained as a result of the following symmetry breaking
\begin{eqnarray}
 & & Fm{\bar 3}m: X_5^+ \nonumber \\
 & & ( \langle S_3(\vec{q}^X_x) \rangle=\langle S_1(\vec{q}^X_y) \rangle
 =\langle S_2(\vec{q}^X_z) \rangle=0 ,
 \nonumber \\
 & & \langle S_2(\vec{q}^X_x) \rangle=\langle S_3(\vec{q}^X_y) \rangle
 =\langle S_1(\vec{q}^X_z) \rangle = \rho_2 \sqrt{N} \neq  0 ) \nonumber \\
 & & \rightarrow Pa{\bar 3} \quad (Z=4) 
 \quad \mbox{domain II}  . \label{Pa3'}
\end{eqnarray}
Previously the symmetry change $Fm{\bar 3}m \rightarrow Pa{\bar 3}$
has been considered in a number of publications, 
Refs.~\onlinecite{Zie,Mic1,NM1,NM2}. The $X_5^+$ active representation
is defined by the star $\{ \vec{k}_{10} \}$ and the loaded two-dimensional 
representation $\hat{\tau}^9$ in Kovalev's notation. \cite{Kov}

Although the two structures look similar there is one very important
difference between them. {\it The $Pa{\bar 3}$ ordering corresponds to
an effective attraction of the quadrupoles, while the $Pn{\bar 3}m$
ordering results in a repulsive interaction between them,}
see Appendix A for details.
This is the main reason why the $Pa{\bar 3}$ structure 
(but not $Pn{\bar 3}m$)
is found in many molecular solids such as
H$_2$, N$_2$,\cite{Scott} NaO$_2$,\cite{Zie} and 
C$_{60}$.\cite{Sach1,Dav,Sach2,Hei2,Mic1} 
Here we speak about the direct bilinear electronic quadrupole-quadrupole
interaction. 
However, in case of NpO$_2$ there are indirect 
(superexchange \cite{super}) 
interactions via oxygen which may lead to an effective
attraction and therefore, the $Pn{\bar 3}m$ structure can not be
ruled out.
Nevertheless, the $Pa{\bar 3}$ space group is a good candidate for the 
symmetry of the ordered phase of NpO$_2$.
Provided that the superexchange interaction remains the same
for two structures, the $Pa{\bar 3}$ symmetry becomes preferable.
It is interesting to notice that at first the symmetry of the ordered
phase of the pristine C$_{60}$ molecular crystal was identified
as $Pn{\bar 3}$.\cite{Hei2} 
Only later it was corrected and proved to be
$Pa{\bar 3}$. \cite{Sach1,Dav,Sach2}

Now we will study how the $Pn{\bar 3}m$ and $Pa{\bar 3}$ structures
manifest themselves in resonant X-ray scattering experiments. \cite{Tem1,Pai}
The tensor of scattering on the quadrupolar density of a neptunium atom 
in a cubic lattice is given by \cite{Dmi1,Dmi2,Tem2}
\begin{eqnarray}
    \hat{f}_n = \tilde{f} \begin{pmatrix} 0 & \rho_1 & \rho_2 \\
    \rho_1 & 0 & \rho_3 \\ \rho_2 & \rho_3 & 0  \end{pmatrix} ,
\label{3f.1}
\end{eqnarray}
where $\rho_i=\pm1$. We have the following 
correspondence \cite{Dmi2,Tem2}
between the functions ${\cal S}_n$, Eq.~(\ref{3c.1a}-d), and 
the scattering tensors: the function ${\cal S}_a$ corresponds
to $\hat{f}_a$, where $\rho_1=\rho_2=\rho_3=1$, the function
${\cal S}_b$ to $\hat{f}_b$ with $\rho_1=1$, $\rho_2=\rho_3=-1$;
the function ${\cal S}_c$ to $\hat{f}_c$ with $\rho_1=\rho_2=-1$, 
$\rho_3=1$; and ${\cal S}_d$ to $\hat{f}_d$ with $\rho_1=\rho_3=-1$, 
$\rho_2=1$. The scattering tensor structure amplitude is found as
\begin{eqnarray}
    {\cal F}(h,k,l) = \sum_{n=1}^N \hat{f}_n \, 
    e^{i\vec{K} \cdot \vec{X}(n)} ,
\label{3f.2}
\end{eqnarray}
where $\vec{K}=(2\pi/a)(h,k,l)$ stands for the vectors of the 
reciprocal lattice and $a$ is the cubic lattice constant. 
In performing the summation in (\ref{3f.2})
we distinguish four contributions from four sublattices
$\{ n_p \}$, $p=1-4$. We are mainly interested in superstructure
Bragg reflections. We obtain then that in general
\begin{eqnarray}
    {\cal F}(h,k,l) = \tilde{F} {\cal M} ,
\label{3f.3}
\end{eqnarray}
where $\tilde{F}=\tilde{f}N$ and the matrix ${\cal M}$ is
either $A$, $B$, or $C$:
\begin{eqnarray}
    A= \begin{pmatrix} 0 & 0 & 0 \\
    0 & 0 & 1 \\ 0 & 1 & 0  \end{pmatrix} , \quad
    B= \begin{pmatrix} 0 & 0 & 1 \\
    0 & 0 & 0 \\ 1 & 0 & 0  \end{pmatrix} , \quad
    C= \begin{pmatrix} 0 & 1 & 0 \\
    1 & 0 & 0 \\ 0 & 0 & 0  \end{pmatrix} . \nonumber \\
\label{3f.4}
\end{eqnarray}
In Table \ref{table0a} we quote which of the matrices ($A,B,C$) occurs for
every particular case of $(h,k,l)$ for the $Pn{\bar 3}m$
and $Pa{\bar 3}$ structures.
%
\begin{table} 
\caption{
Tensor ${\cal F}(h,k,l)=\tilde{F} {\cal M}$ for the
superstructure Bragg reflections on quadrupolar densities of Np.
${\cal M}=A,B$ or $C$, Eq.~(\ref{3f.4}), I $(ii)$ and II $(iii)$ refer to
two domains of the $Pa{\bar 3}$ structure described in the text, $0$
is the zero matrix.
\label{table0a}     } 
 
\begin{ruledtabular}
 \begin{tabular}{l  c  c  c} 
Bragg reflection  & $Pn{\bar 3}m$ & \multicolumn{2}{c}{$Pa{\bar 3}$} \\
  $h,k,l:$  &  $(i)$        & I $(ii)$   &  II $(iii)$ \\
\hline
 $h,k,l=2n$   & $0$ & $0$ & $0$ \\
 $h,k,l=2n+1$ & $0$ & $0$ & $0$ \\
 $h=2n+1$, $\;k,l=2n$ & $A$ & $C$ & $B$ \\
 $k=2n+1$, $\;h,l=2n$ & $B$ & $A$ & $C$ \\
 $l=2n+1$, $\;h,k=2n$ & $C$ & $B$ & $A$ \\
 $h,k=2n+1$, $\;l=2n$ & $C$ & $B$ & $A$ \\
 $h,l=2n+1$, $\;k=2n$ & $B$ & $A$ & $C$ \\
 $k,l=2n+1$, $\;h=2n$ & $A$ & $C$ & $B$ 
 \end{tabular} 
\end{ruledtabular}
\end{table} 
We recall that the conditions \cite{Tables} for the
isotropic scattering on the fcc lattice of Np
(from the spherically symmetric densities) are
$h+k,k+l=2n$, that corresponds to two first lines
of Table \ref{table0a}.

Finally, we consider the polarization dependencies for
resonant X-ray scattering experiments. We assume that
a crystal has a flat (001) surface and the azimuthal 
angle $\psi$ is counted from the $x-$axis defined by 
$\vec{e}_x=(1,0,0)$. \cite{Pai} 
The incident beam direction is
given by ($-\cos \Theta \,\cos \psi$, $-\cos \Theta \, \sin \psi$, 
$-\sin \Theta$),
the scattered beam direction by
($-\cos \Theta\,\cos \psi$, $-\cos \Theta \, \sin\psi$, $\sin \Theta$).
We introduce the standard polarization vectors $\vec{e}_i$ ($i=1-3$) 
parallel ($\vec{\pi}_i$, $\vec{\pi}_f$)
or perpendicular ($\vec{\sigma}$) to the scattered plane,  
\begin{subequations}
\begin{eqnarray}
  & &\vec{\sigma}=(-\sin \psi,\; \cos \psi,\; 0) , \label{3f.5a} \\
  & &\vec{\pi}_i=(-\sin \Theta \cos \psi,\; -\sin \Theta \sin \psi,\; 
  \cos \Theta) , \quad \quad \quad \label{3f.5b} \\
  & &\vec{\pi}_f=( \sin \Theta \cos \psi,\;  \sin \Theta \sin \psi,\; 
  \cos \Theta) . \label{3f.5c}
\end{eqnarray}
\end{subequations}
A structure factor amplitude $F(h,k,l)$ is found as \cite{Dmi1,Dmi2,Tem2}
\begin{eqnarray}
   F_{p \rightarrow p'}(h,k,l)=\tilde{F}\, 
   \vec{e}_{p'}{\,}^T  {\cal M} \, \vec{e}_p ,
 \label{3f.6}
\end{eqnarray}
where $T$ stands for transpose and
$p \rightarrow p'$ denotes one of the four polarization channels,
$\sigma \rightarrow \sigma$, $\sigma \rightarrow \pi_f$,
$\pi_i \rightarrow \sigma$, and $\pi_i \rightarrow \pi_f$.
The structure factor amplitudes for the three cases (${\cal M}=A,B,C$)
are quoted in Table \ref{table0b}. 
%
\begin{table} 
\caption{
The structure factor amplitudes $F_{p \rightarrow p'}(h,k,l)/\tilde{F}$
for four polarization channels of RXS and the matrices $A$, $B$ and $C$.
\label{table0b}     } 
 
\begin{ruledtabular}
 \begin{tabular}{l  c  c  c} 
 $p \rightarrow p'$ & $A$ & $B$ & $C$ \\
\hline
 $\sigma \rightarrow \sigma$ & 0                       & 
   0                        & $-\sin(2\psi)$ \\
 $\sigma \rightarrow \pi_f$  & $\cos \Theta \cos \psi$ & 
   $-\cos \Theta \sin \psi$ & $\cos (2\psi)\, \sin \Theta$ \\
 $\pi_i \rightarrow \sigma$  & $\cos \Theta \cos \psi$ & 
   $-\cos \Theta \sin \psi$ & $-\cos (2\psi)\, \sin \Theta$ \\
 $\pi_i \rightarrow \pi_f$   & 0                       & 
   0                        &  $\sin (2\psi)\, \sin^2 \Theta$ 
 \end{tabular} 
\end{ruledtabular}
\end{table} 
The corresponding intensities of the 
superstructure Bragg reflections are found as 
$I_{p \rightarrow p'}(h,k,l)=|F_{p \rightarrow p'}(h,k,l)|^2$.

From Tables \ref{table0a} and \ref{table0b} one can easily 
obtain all necessary dependencies for intensities of
different polarizations. For example, the intensity of
the $(003)$ reflection of $Pn{\bar 3}m$ is exactly the same as
the intensity of the same polarization of the $(300)$ reflection
of domain I $(ii)$ of $Pa{\bar 3}$, and the intensity of 
the $(030)$ reflection of domain II $(iii)$ of $Pa{\bar 3}$. 
Furthermore, the $(013)$ reflection of domain I $(ii)$,
and the $(103)$ reflection of domain II $(iii)$ of $Pa{\bar 3}$
also have the same intensity, and so on.
Since both symmetries are very similar, we believe that special care 
should be taken to discriminate between the $Pn{\bar 3}m$ and $Pa{\bar 3}$
structures.

Tables \ref{table0a} and \ref{table0b} are also useful in
considering the contributions from the domains of the same group.
We start with the $Pn{\bar 3}m$ structure.
The four domains differ by the orientation
of the $Y_2^0(\Omega')$ quadrupolar function at $\vec{n}=0$.
This function may be chosen to align along four main cube diagonals,
which are connected with each other through rotations 
by $\pi/2$ about the $z-$axis.
Hence,
all four domains are obtained from the first, $(i)$, by applying
three consecutive rotations by $\pi/2$ about the $z-$axis.
The effect of the domains in RXS experiments is equivalent
to a superposition of four structure amplitudes, Table \ref{table0b},
with angles $\psi$, $\psi+\pi/2$, $\psi+\pi$, and $\psi+3\pi/2$.
(Here we suggest that the domains produce the coherent
scattering.)
We label these domains by indices $d1-d4$, and introduce their
populations, $P_{d1}-P_{d4}$. The population of a domain $d$
is defined as $P_d=N_d/N$, where $N_d$ is the total number of
neptunium atoms in the domain.
Then we obtain from Eq.\ (\ref{3f.6}) that the structure amplitude 
for the $\sigma \rightarrow \sigma$ and $\sigma \rightarrow \pi_f$
channels of the $(003)$ reflection is given by
\begin{subequations}
\begin{eqnarray}
   F_{\sigma \rightarrow \sigma}(003) &=&
   -\tilde{F}\, \tilde{P}\, \sin(2\psi) , 
 \label{3f.7a} \\
   F_{\sigma \rightarrow \pi_f}(003) &=& \tilde{F}\, \tilde{P}\,
   \cos(2\psi) \sin \Theta , 
  \label{3f.7b}   
\end{eqnarray}
where
\begin{eqnarray}
 \tilde{P}=
   \left( P_{d1} - P_{d2} + P_{d3} - P_{d4} \right) .
  \label{3f.7c}   
\end{eqnarray}
\end{subequations}
If $P_{d1} + P_{d3} = P_{d2} + P_{d4}$, $\tilde{P}=0$
and the $(003)$ Bragg reflection is suppressed.
Otherwise, the $\psi-$ and $\Theta-$dependencies in (\ref{3f.7a},b)
are exactly the same as for a single domain, Tables \ref{table0a},
\ref{table0b}.
For the $(300)$ reflection of $Pn{\bar 3}m$ one obtains that
\begin{subequations}
\begin{eqnarray}
   F_{\sigma \rightarrow \pi_f}(300)=\tilde{F}\, P' \,
   \cos \Theta \cos (\psi+\psi_0) ,  \label{3f.8a}
\end{eqnarray}
where
\begin{eqnarray}
 & & \psi_0 = \arccos ([P_{d1} - P_{d3}]/P')  ,
 \label{3f.8b} \\
 & & P'=\sqrt{[P_{d1} - P_{d3}]^2 + [P_{d2} - P_{d4}]^2} .
 \label{3f.8c}
\end{eqnarray}
\end{subequations}
Comparing this result with the expression for a single domain, 
we observe that the main effect
is the phase shift $\psi_0$ given by Eq.~(\ref{3f.8b}).
The condition for suppression of the reflection is
$P_{d1}=P_{d3}$ and $P_{d2}=P_{d4}$.
The polarization dependencies of the other reflections of
$Pn{\bar 3}m$ can be figured out
analogously.

There are eight domains of $Pa{\bar 3}$ structure. 
Earlier we have already considered the two basic variants
of $Pa{\bar 3}$: I $(ii)$, and II $(iii)$. 
The others are obtained by rotating these
two variants by the angles $\pi/2$, $\pi$, and $3\pi/2$ about the
$z-$axis. Applying the rotations to I and II, we arrive at two
series. We label the population of the series of domains by 
the indices $P^I_{d1}-P^I_{d4}$, and $P^{II}_{d1}-P^{II}_{d4}$,
respectively. 
The two series result in two distinct scattering matrices,
for example, for $(003)$ they are $B$ and $A$, for $(300)$
$C$ and $B$, etc, see Table \ref{table0a}.
The domain pattern produces
a superposition of 8 terms (amplitudes). 
Each term corresponds to one of the two matrices and to one of
the four azimuthal angles: $\psi$, $\psi+\pi/2$, $\psi+\pi$,
and $\psi+3\pi/2$. The relevant expressions are obtained
in the same way which we have used to derive Eqs.~(\ref{3f.7a}-c) and 
(\ref{3f.8a}-c).
For example,
\begin{subequations}
\begin{eqnarray}
   F_{\sigma \rightarrow \pi_f}(003)&=&0 ,
\label{3f.9a} \\
   F_{\sigma \rightarrow \pi_f}(003)&=&
   \tilde{F}\,\cos \Theta ( -P'{}^I \,
   \sin (\psi+\psi_0^I) \nonumber \\
   & &+P'{}^{II} \cos (\psi+\psi_0^{II}) )  ,  
\label{3f.9b}
\end{eqnarray}
\end{subequations}
The latter expression can be transformed to a shifted $sin-$ or
$cos-$ like functions, i.e. 
$F_{\sigma \rightarrow \pi_f}(003) \sim \cos \Theta \sin(\psi +\psi'_0)$,
where $\psi'_0$ is a phase shift depending on the domain pattern
of $Pa{\bar 3}$. $P'{}^I$, $P'{}^{II}$, and $\psi_0^I$, $\psi_0^{II}$ 
are given by Eq.~(\ref{3f.8c}) and (\ref{3f.8b}), where $P_{di}$ ($i=1-4$) 
are replaced by $P_{di}^I$ or $P_{di}^{II}$. 
For the $(300)$ reflection of $Pa{\bar 3}$ we find
\begin{subequations}
\begin{eqnarray}
  & & F_{\sigma \rightarrow \sigma}(300) = -\tilde{F}\,\tilde{P}^I
   \sin(2\psi) , 
\label{3f.10a} \\
  & & F_{\sigma \rightarrow \pi_f}(300) =
   \tilde{F}\,\tilde{P}^I \cos(2\psi) \sin \Theta \nonumber \\
  & & \quad -\tilde{F}\,P'{}^{II} \sin(\psi+\psi_0^{II}) \cos \Theta  . 
  \quad 
\label{3f.10b}
\end{eqnarray}
\end{subequations}
Here again $\tilde{P}^I$ is given by (\ref{3f.7c}) for $P_{di}^I$, 
$i=1-4$.
As before, the superstructure Bragg reflections are suppressed
in the case of equal population of 8 domains in the crystal.

\section {Configuration interaction calculation} 

\label{sec:met} 

In this section we describe in detail how we construct
the basis consisting of many determinantal wave functions 
for a many electron system and how we calculate the relevant
matrix elements.
First, we perform a band structure calculation in order to
determine the conduction electron charge distribution in
the muffin-tin (MT) sphere around Np.
Subsequently, we apply the configuration interaction method
to treat the many electron system consisting of the localized
5$f^3$ electrons and the conduction electron in the MT sphere.

\subsection{Electron band structure calculation}

We have started by performing an electron band structure
calculation of NpO$_2$ using our linear augmented plane wave
(LAPW) code. \cite{Ion} The calculation 
has been done assuming the muffin-tin (MT) shape of the one-electron
potential and the Barth-Hedin expression for exchange, \cite{Bar}
which is a variant of the local density approximation (LDA).
The equal MT radii 
$R_{MT}^{Np}=$2.2206 and $R_{MT}^{O}=$2.2206 in atomic units (a.u.)
were chosen for Np and O,
with the cubic lattice constant $a=10.2567$ a.u.
(or 5.4276 {\AA}). \cite{Man}
The MT potential and density of Np and O have been obtained
self-consistently using a LAPW basis of $\sim$300 functions
on a 20-point mesh of the irreducible part of the Brillouin zone.
The three $5f$ electrons of Np were treated as core states,
which adjust self-consistently to the conduction electron density.

As a result of the calculation we obtained that NpO$_2$ is an
insulator, with the energy gap $\triangle E=0.789$ eV.
The width of the occupied electron band below the Fermi level
is $E_T-E_B=5.953$ eV.
The spin-orbit splitting between $5f_{7/2}$ and
$5f_{5/2}$  one-electron states is $\triangle_f=$0.983 eV.
The main goal of the calculation however was the electron charge
density distribution inside the neptunium MT-sphere.
The calculated partial charges of different angular symmetry
($l=0-3$) are quoted in Table \ref{table1}.
%
\begin{table} 
\caption{
Angular-momentum-decomposed 
(partial) electronic charges $Q_l^A$ and
total charges $Q^A$
inside neptunium and oxygen MT spheres and 
in the interstitial
region (LAPW calculations, see Ref.~\onlinecite{Sin} for 
details and definitions); $Q^{Np}=-(2Q^O+Q^i)$.
\label{table1}     } 
 
\begin{ruledtabular}
 \begin{tabular}{c | r c c c} 
  $A$ & Np & O & interstices/per unit \\
\hline
$Q_s^A$    &  0.041$e$ & 0.027$e$ &  $-$  \\
$Q_p^A$    &  0.401$e$ & 4.313$e$ &  $-$ \\
$Q_d^A$    &  0.261$e$ & 0.046$e$ &  $-$ \\
$Q_f^A$    &  2.185$e$ & 0.011$e$ &  $-$ \\
\hline
$Q^A$ & +4.108$|e|$ & -0.284$|e|$ & -3.540$|e|$ 
 \end{tabular} 
\end{ruledtabular}
\end{table} 
An important result is that on average there is 
approximately one conduction
electron present inside the Np MT-sphere.
Therefore, the localized $5f^3$ configuration of Np can
not be considered separately from this valence electron,
which can be of $7s$, $7p$ or $6d$ type.
The instantaneous configuration
at the neptunium site becomes $7s5f^3$, $7p5f^3$ or $6d5f^3$.

In all cases this additional electron experiences a
strong Coulomb repulsion with the three localized partners.
This interaction is not fully accounted for by the
band structure calculations 
because it requires a multideterminant treatment 
or configuration interaction (CI). \cite{NM3}
Therefore, we have to follow a different route and
below we study the electron spectrum using
the multipole expansion of electronic densities.

\subsection{Many electron basis states}

Our method of multipole expansion of the Coulomb interaction has been
used before. \cite{NM1,NM2,NM3,NM4}
Here we formulate it in detail for the $sf^3$ configuration
following Refs.\ \onlinecite{NM3,NM4}.
In Sec.~IV we deal with the $5f^3$ configuration which is easily
obtained from $sf^3$ by omitting one $s$ electron.
In Sec.~V we consider $7s5f^3$, $7p5f^3$ and $6d5f^3$ configurations.
For the latter two cases we will describe the important differences
with $sf^3$.

We start by considering a face centered cubic (fcc) crystal of $N$ Np atoms.
Each atomic site possesses one 7$s$ and three 5$f$ electrons.
The position vector of an electron near a crystal lattice site $\vec{n}$
is given by
\begin{eqnarray}
 \vec{R}(\vec{n}) = \vec{X}(\vec{n})+\vec{r}(\vec{n}) .
\label{2.1} 
\end{eqnarray}
Here $\vec{X}(\vec{n})$ is the lattice vector which specifies
the centers of the atoms (Np-nuclei) on a rigid fcc lattice.
The radius vector $\vec{r}(\vec{n})$ is given in polar coordinates
by $(r(\vec{n}),\Omega(\vec{n}))$, where $r$ is the length and
$\Omega=(\Theta,\phi)$ stands for the polar angles.
We label the basis ket-vectors at the lattice site $\vec{n}$ 
by a single index $I$ or, alternatively, by four
one electron indices $(i^f_1,i^f_2,i^f_3;\,i^s)$:
\begin{eqnarray}
   | I \rangle_{\vec{n}} = 
   | i^f_1,i^f_2,i^f_3;\,i^s \rangle_{\vec{n}} .
\label{2.2} 
\end{eqnarray}
The index $i^f=(m^f,s^f_z)$ stands for the orbital ($m^f=1-7$) and 
spin projection ($s_z=\pm1/2$) quantum numbers of 
one $f$ electron. Therefore, there are 14 states which we label by
$i^f=1-14$. Two states of the $s$ electron are labeled by $i^s=1,2$. 
The many electron basis wave functions are  
\begin{eqnarray}
  \langle \vec{r}_1, \vec{r}_2, \vec{r}_3, \vec{r}_4  | 
  I \rangle_{\vec{n}} = \frac{1}{\sqrt{N_a}} \sum_a\; (-1)^{P(a)}
  \nonumber \\
  \times \prod_{t=1}^3 \,  
  \langle \vec{r}_t | i^f_t \rangle_{\vec{n}} \cdot
  \langle \vec{r}_4 | i^s \rangle_{\vec{n}} , 
\label{2.3} 
\end{eqnarray}
where $a$ stands for a permutation of four electrons, 
the factor $(-1)^P$ takes into account the parity of the permutation, 
$N_a$ is the number of the permutations, and
\begin{subequations}
\begin{eqnarray}
   & &\langle \vec{r} | i^f \rangle_{\vec{n}} = 
   {\cal R}_f(r(\vec{n})) \langle \hat{n} | i^f \rangle , \label{2.4a} \\ 
   & &\langle \vec{r}\,' | i^s \rangle_{\vec{n}} = 
   {\cal R}_s(r'(\vec{n})) \langle \hat{n}' | i^s \rangle . \label{2.4b}  
\end{eqnarray}
\end{subequations}
Here ${\cal R}_f$ and ${\cal R}_s$ are the radial components of the 
5$f$ and 7$s$ electrons, respectively; 
$\hat{n}$ stands for $\Omega(\vec{n})$. 
The $5f$ spin-orbitals can be written as
\begin{eqnarray}
 & &\langle \hat{n} | i^f \rangle=
 \langle \hat{n} | m_f \rangle\, u_s(s_z(f)) , 
 \label{2.5}
\end{eqnarray}
Here $u_s$ is the spin function ($s=\pm$). The $f-$orbital parts,
$\langle \hat{n} | m_f \rangle$, 
are expressed in terms of spherical harmonics 
$Y_l^m(\Omega)=\langle \hat{n}| l,m \rangle$. 
We find it convenient
to work with real spherical harmonics \cite{Bra} $Y_l^{\tau}$,
where $\tau=0$, $(m,c)$ or $(m,s)$.

The order of indices in (\ref{2.2}) is important.
For example, as follows from the dynamical
equivalence of the electrons the state $| i^f_1,i^f_2,i^s,i^f_3 \rangle$
can be reduced to $| i^f_1,i^f_2,i^f_3;\,i^s \rangle$
by permuting the third and the fourth electrons, i.e.
\begin{eqnarray}
 | i^f_1,i^f_2,i^s,i^f_3 \rangle=-| i^f_1,i^f_2,i^f_3;\,i^s \rangle,
\label{2.7} 
\end{eqnarray}
and so on. To describe the same quantum state we will use
the basis vectors (\ref{2.2}) and apply the permutation
law (\ref{2.7}) when needed.
Alternatively, one can use the corresponding Slater
determinants for
the four electron wave functions, Eq.\ (\ref{2.3}).
However, the permutation relations of the type of Eq.~(\ref{2.7})
are more efficient for our purposes. 
Excluding equivalent states, we find only
$(14\cdot 13 \cdot 12/3!)\times 2=728$ independent functions,
or determinants, for $7s5f^3$.
(These are 364, 2184 and 3640 for $5f^3$, $7p5f^3$, $6d5f^3$, 
respectively.)
Notice, that every basis wave function is
in fact a Slater determinant, Eq.\ (\ref{2.3}).

\subsection{Multipole repulsion between electrons}

Now we take into account the Coulomb intrasite and intersite 
repulsion by expanding the interaction in multipole series.
As was discussed in Ref.\ \onlinecite{NM3}, these interactions
are treated exactly in the chosen quantum space ($7s5f^3$).

The Coulomb interaction between two electrons at
sites $\vec{n}$ and $\vec{n}'$ is given by
\begin{eqnarray}
  V(\vec{R}(\vec{n}),\vec{R}'(\vec{n}'))=
  {\frac{1}{|\vec{R}(\vec{n})-\vec{R}'(\vec{n}')|}} .
\label{2.8'} 
\end{eqnarray}
The multipole expansion 
in terms of site symmetry adapted functions (SAF's) \cite{Bra} is 
\begin{eqnarray}
 & & V(\vec{R}(\vec{n}),\vec{R}'(\vec{n}'))=
  \sum_{\Lambda \Lambda'} 
  v_{\Lambda \Lambda'}(\vec{n},\vec{n}';\,r,r')\,
  S_{\Lambda}(\hat{n})\, S_{\Lambda'}(\hat{n}'), \nonumber \\
 & & \label{2.8} 
\end{eqnarray}
where
\begin{eqnarray}
 & & v_{\Lambda \Lambda'}(\vec{n},\vec{n}';\,r,r')\,
   =  \int \! d\Omega(\vec{n}) \int \! d\Omega'(\vec{n}')\,
      {\frac{  S_{\Lambda}(\hat{n})\, S_{\Lambda'}(\hat{n}')}
     {|\vec{R}(\vec{n})-\vec{R}'(\vec{n}')|}} . \nonumber   \\
 & & \label{2.9} 
\end{eqnarray}
The SAF's are linear combinations of spherical
harmonics and transform as irreducible representations
of a site point group, Ref.\ \onlinecite{Bra}.
The index $\Lambda$ stands for $(l,\tau)$, with $\tau=(\Gamma,\mu,k)$.
Here $l$ accounts for the angular dependence of the
multipolar expansion, $\Gamma$ denotes an
irreducible representation (in the present case the 
group is $O_h$),
$\mu$ labels the representations that occur more than once and
$k$ denotes the rows of a given representation.

The intrasite case corresponds to $\vec{n}=\vec{n}'$. 
The interaction function 
$v_{\Lambda \Lambda'}(r,r') \equiv  
v_{\Lambda \Lambda'}(\vec{n}=\vec{n}';\,r,r')$
then becomes particular simple,
\begin{eqnarray}
  v_{\Lambda \Lambda'}(r,r') \,
  = \left( {\frac {r^l_< }{r^{(l+1)}_> }} \right)
  {\frac {4\pi}{2l+1}} 
  \delta_{\Lambda \Lambda'} , 
\label{2.10} 
\end{eqnarray}
where $r_>=max(r,r')$, $r_<=min(r,r')$ and 
$\delta_{\Lambda \Lambda'}=\delta_{\tau \tau'} \delta_{l l'}$.
The last expression is also site independent.

There is no simple analytical expression for the intersite case,
$\vec{n} \neq \vec{n}'$. \cite{Hei} The intersite multipole
interactions are anisotropic and for practical purposes
it is important to use
the following dependence \cite{Hei}
\begin{eqnarray}
 v_{\Lambda \Lambda'}(\vec{n},\vec{n}';\,r,r') 
 \sim \frac{(r)^l (r')^{l'}}{|\vec{X}(\vec{n})-\vec{X}(\vec{n}')|^{l+l'+1}} .
 \label{2.11} 
\end{eqnarray}

\subsection{Intrasite matrix elements}

For the Coulomb interaction between four electrons {\it on a same site}
$\vec{n}$ we have a sum of six two-body terms,
\begin{eqnarray}
 V^{(4)}= \frac{1}{2}
 \sum_{t=1}^4 \sum_{p (\neq t)=1}^4 V(\vec{r}_t,\vec{r}_p) ,
\label{2.14n} 
\end{eqnarray}
where each term is given by the multipole expansion (\ref{2.8}).
In order to calculate the matrix elements of $V^{(4)}$, 
$\langle i^f_1, i^f_2, i^f_3;\, i^s  |V^{(4)}| 
j^f_1, j^f_2, j^f_3;\, j^s \rangle$,
one has to classify the electronic transitions.
Following Ref.\ \onlinecite{NM4} where the energy terms
of molecular ions C$_{60}^{m\pm}$, $m=2-5$, were calculated,
we consider four possibilities 
for the fourth $s-$electron:
(1) $i^s \rightarrow j^s$, (2) $i^s \rightarrow j^f_3$, 
(3) $i^s \rightarrow j^f_2$, and (4) $i^s \rightarrow j^f_1$,
which we label by the index $a_4=1-4$. The $a_4=2$ and $a_4=4$ 
transitions involve odd number of transpositions among 
$j^f_1,j^f_2,j^f_3;j^s$, and
the parity is $P(a_4=2)=P(a_4=4)=-1$. For two other
transitions the number of transpositions is even and 
$P(a_4=1)=P(a_4=3)=1$.
After this we are left with only
three $j$-states, which we label as $j'_1$, $j'_2$, and $j'_3$.
For the next electron, $i^f_3$, we can consider three possibilities
($i^f_3 \rightarrow j'_3$, $i^f_3 \rightarrow j'_2$, 
$i^f_3 \rightarrow j'_1$) which we label by the index $a_3=1-3$.
In this way we continue until we exhaust all four electrons. 
As a result, each
subcase (or electron transition) is classified by the
three index label $a \equiv (a_4,a_3,a_2)$, 
and its parity is $P(a)=P(a_4)P(a_3)P(a_2)$. 
Mathematically, we reduce
a permutation of four electrons to a product of transpositions. 
The matrix element $\langle I|V^{(4)}|J \rangle$ is found as \cite{QC49}
\begin{eqnarray}
 \langle I |V^{(4)}| J \rangle = \sum_{a} P(a)
 \langle I  | V^{(4)} | J \rangle^{(a)} ,
\label{2.14} 
\end{eqnarray}
where $\sum_a=\sum_{a_4=1}^4 \sum_{a_3=1}^3 \sum_{a_2=1}^2$,
and 
\begin{eqnarray}
 \langle I  | V^{(4)} | J \rangle^{(a)} &=& 
 \sum_{l,\tau} 
  v_{l}^{s,j_{a4}-f,j_{a3}} \, 
  c_{l,\tau}(i^s j_{a4})\, c_{l,\tau}(i^f_3 j_{a3})\, 
 \nonumber \\
  & & \times  \delta(i^f_2,j_{a2}) \delta(i^f_1,j_{a1}) 
  +\,p.i. 
\label{2.15n} 
\end{eqnarray}
Here $p.i.$ stands for the other pair Coulomb interactions,
Eq.\ (\ref{2.14n}). (The explicitly written term in Eq.\ (\ref{2.15n})
corresponds to the interaction between fourth and third electron.)
The elements $c_{\Lambda}(ij) \equiv c_{l, \tau}(ij)$ are defined by
\begin{eqnarray}
  c_{\Lambda}(i j) =  
  \int \! d\Omega \, \langle i |\hat{n}\rangle  
  S_{\Lambda}(\hat{n}) \langle \hat{n}|j \rangle .
\label{2.15} 
\end{eqnarray}
For the $7s5f^3$ configuration there are three types of these
coefficients. For the $s-s$ transition it is only one integral
$\langle s |Y_0^0| s \rangle=1/\sqrt{4\pi}$, which is not zero.
For the $f-f$ transitions and real spherical harmonics, \cite{Bra}
the coefficients $c_{l,\tau}(i^f\,j^f)$ were tabulated in Ref.~\onlinecite{NM1}.
Finally, there are $f-s$ and $s-f$ transitions which require
the evaluation of $c_{l,\tau}(i^f\,j^s)$. From the orthogonality
of spherical harmonics we find that 
\begin{eqnarray}
 \langle 0,0 |Y_3^{\tau}| 3, \tau \rangle = \frac{1}{\sqrt{4\pi}} ,
\label{2.16} 
\end{eqnarray}
where $\tau=0$, $(m,c)$, $(m,s)$, $m=1-3$, and zero otherwise.
The matrix quantities (\ref{2.15}) were first introduced
by Condon and Shortley for the description of atomic spectra, \cite{CS}
but they are also at the center of the calculation of the crystal
electric field effects. \cite{NM1}

In Eq.~(\ref{2.15n}) $v_l^{s,j_{a4}-f,j_{a3}}$ stands for a radial average.
The general expression is
\begin{eqnarray}
  v_{l}^{a,b-c,e} =
  \int \! dr\, r^2 \int \! dr'\, {r'}^2\,  {\cal R}_a(r){\cal R}_b(r)\,
  \nonumber \\ 
   {\cal R}_c(r'){\cal R}_e(r')\,
  v_l(r,r') , 
\label{2.17} 
\end{eqnarray}
where ${\cal R}_a$, ${\cal R}_b$, ${\cal R}_c$ and 
${\cal R}_e$ are radial components, and 
$v_l(r,r')=v_{\Lambda \Lambda}(r,r')$
is given by Eq.~(\ref{2.10}).
For the $7s5f^3$ configuration the indices $a,b,c,e$ refer
either to $7s$ or to $5f$ electron radial components.
There are only two types of radial integrals,
corresponding to nontrivial multipolar terms ($l \neq 0$)
of $7s5f^3$, which are
\begin{subequations}
\begin{eqnarray}
  v_l^{ff-ff} &=&
  \int \! dr\, r^2 \int \! dr'\, {r'}^2\,  
  {\cal R}_f^2(r)\, {\cal R}_f^2(r')\,
  v_l(r,r') , \quad  \quad \quad \label{2.18a}  \\ 
  v_{l'}^{fs-sf} &=&
  \int \! dr\, r^2 \int \! dr'\, {r'}^2\,  
  {\cal R}_s(r){\cal R}_f(r)\,  \nonumber \\
  & & {\cal R}_f(r'){\cal R}_s(r')\,
  v_l(r,r') . \quad  
  \label{2.18b}
\end{eqnarray}
\end{subequations}
Here $l=2,4,6$ and $l'=3$, as follows from the selection rules
for $c_{\Lambda}(ff)$ and $c_{\Lambda'}(fs)$.
The radial integrals $v_l^{ff-ff}$ and $v_{l'}^{fs-sf}$ are proportional
to the quantities $F_l$ and $G_{l'}$ introduced by Condon and Shortley
in Ref.~\onlinecite{CS}.
It is important to notice that even the $7s$ electron with
the trivial dependence of its angular part is strongly coupled
to the three $5f$ electrons through $f-s$ and $s-f$ transitions.

The classification scheme for electronic transitions
which we have introduced here
is very useful for handling the single particle interactions as well.
The main difference is that now the interaction occurs to
a single electron while the rest of them produce Kronecker factors,
Ref.~\onlinecite{NM4}. This group of interactions includes
the spin-orbit coupling $H_{so}$, the crystal electric field
$V_{CF}$ and the mean field $V_{MF}$. The latter two interactions
are dealt with in Secs.~IV and V.  The spin-orbit coupling
is
\begin{eqnarray}
  H_{so}=\sum_{i} V_{so}(i) ,
  \label{2.19'a}
\end{eqnarray}
where the sum runs over all electrons, $V_{so}$ being 
the corresponding one-electron spin-orbit operator. 
The $s-$electron does not experience  
the spin-orbit coupling and in the $7s5f^3$ case the summation
includes only three $5f$ terms $V_{so}(i^f)$, $i^f=1-3$, where
\begin{eqnarray}
   V_{so}(i^f)=\zeta_f {\vec L}(i^f) \vec{S}(i^f) .
  \label{2.19'b}
\end{eqnarray}
Here ${\vec L}(i^f)$ and $\vec{S}(i^f)$ are the one-electron operators
of orbital and spin momentum, $\zeta_f$ is the constant
of the spin-orbit coupling. The full intrasite Hamiltonian
is given by $H_{intra}=V^{(4)}+H_{so}$. It describes
the $7s5f^3$ configuration of a free neptunium ion.
Since the present method is not based on perturbation
theory it extends the classical calculations of Condon
and Shortley. \cite{CS}

\subsection{Intersite matrix elements}

We start with expression (\ref{2.8}) and write it in the space 
of many electron basis vectors $| I \rangle$, Eq.~(\ref{2.2}).
Carrying out the angular integrations $d\Omega(\vec{n})$,
$d\Omega'(\vec{n})$, $d\Omega(\vec{n}')$, $d\Omega'(\vec{n}')$,
we obtain
\begin{eqnarray}
 & & \langle I |_{\vec{n}} \langle I' |_{\vec{n}'}
 V(\vec{R}(\vec{n}),\vec{R}'(\vec{n}'))
  |J' \rangle _{\vec{n}'}| J \rangle _{\vec{n}} = 
     \frac{1}{N_a}
     \nonumber \\ 
  & \times & \sum_{a(\vec{n})} \sum_{a'(\vec{n}')}
  P(a_{\vec{n}}) P(a'_{\vec{n}'})\,
  \sum_{\alpha=1}^4 \sum_{\alpha'=1}^4 \sum_{\Lambda \Lambda'}
  v_{\Lambda}^{\alpha \alpha}{\,}_{\Lambda'}^{\alpha' \alpha'}
  (\vec{n}-\vec{n}')
     \nonumber \\
 & \times & 
  \left\{ c_{\Lambda}(i_{\alpha} j^a_{\alpha})
  \prod_{\beta=1}^3 \delta(i_{\beta} j^a_{\beta}) \right\} 
  \left\{ c_{\Lambda'}(i'_{\alpha'} j'{}^{a'}_{\alpha'}) 
  \prod_{\beta'=1}^3 \delta(i'_{\beta'} j'{}^{a'}_{\beta'})  
  \right\} .   \nonumber \\
\label{2.19} 
\end{eqnarray}
Here the sum over $a$ means the summation over all permutations of
indices $j^f_1,j^f_2,j^f_3,j^s$ at site $\vec{n}$ transforming them
to indices $j^a_{\alpha}$ ($\alpha=1-4$). Analogously, the sum
over $a'$ implies the summation over all permutations of ${j'}^f_1,{j'}^f_2,{j'}^f_3,{j'}^s$ 
at site $\vec{n}'$ transforming them to $j'{}^{a'}_{\alpha}$ ($\alpha=1-4$).
$P(a)$ and $P(a')$ stand for the parities
of the permutations.
Indices $\alpha$ and $\alpha'$ indicate which electron at site $\vec{n}$
interacts with which electron at site $\vec{n}'$. The other electrons
labeled by $\beta=1-3$ at site $\vec{n}$ and by $\beta'=1-3$ at site 
$\vec{n}'$ do not contribute to the interaction and produce the
Kronecker delta symbols. 
The coefficients $c_{\Lambda}$ are defined by 
Eq.\ (\ref{2.15}), and the intersite ($\vec{n} \neq \vec{n}'$) 
interaction element 
$v_{\Lambda}^{\alpha \alpha}{\,}_{\Lambda'}^{\alpha' \alpha'}$
is given by
\begin{eqnarray}
  v_{\Lambda}^{\alpha \alpha}{\,}_{\Lambda'}^{\alpha' \alpha'}(\vec{n}-\vec{n}')
   =
  \int \! dr\, r^2 \int \! dr'\, {r'}^2\,  \nonumber \\
   \times  {\cal R}_{\alpha}^2(r)\, {\cal R}_{\alpha'}^2(r')\,
  v_{\Lambda \Lambda'} (\vec{n},\vec{n}';\, r,r') .
\label{2.20} 
\end{eqnarray}
For the four electron space of $7s5f^3$ only even $l$ and $l'$ 
(in $\Lambda,\Lambda'$) are retained in
Eqs.~(\ref{2.19}) and (\ref{2.20}), and $l,l'=0,2,4,6$.
Two very important examples of intersite Coulomb interactions,
namely, the crystal electric field and the mean field will be considered
in sections IV and~V.

\section {Crystal and mean field of the $5f^3$ 
neptunium configuration} 

\label{sec:f3} 

In this section we study the model where we assume
that there are only three localized $5f$ electrons
at each neptunium site.
Although we believe that the model is not adequate
for a realistic description of NpO$_2$, especially
in the part concerning the loss of the magnetic moments
in the ordered phase, it is nevertheless very instructive
to consider it in detail. The $5f^3$ configuration being relatively simple
offers an opportunity to be studied thoroughly
and to understand the interplay between the disordered
and ordered phases, or between the crystal and 
quadrupolar mean field.
On the other hand, the configurations $7p5f^3$ and $6d5f^3$
involve too many basis states and consume to much time
to be processed self-consistently for any temperature.

\subsection{Free ion electron energy spectrum}

The basis states for the $5f^3$ configuration are given by
\begin{eqnarray}
  | I \rangle = | i^f_1, i^f_2, i^f_3 \rangle,
\label{3.3} 
\end{eqnarray}
where as before, $i^f=1-14$. The total number of basis
vectors is $14 \times 13 \times 12/3!=364$.
We treat the $5f^3$ configuration in the way 
which was specified in Sec.~III.

There are only $f-f$ transitions described by
four radial integrals (\ref{2.18a}): $v^{ff-ff}_l$, 
$l=0,2,4,6$.
The others are zero due to the selection rules imposed
by the coefficients $c_{\Lambda}(ij)$, Eq.~(\ref{2.15}).
The radial integral $v^{ff-ff}_0$ (Hubbard $U$) is not important
here since it does not result in term splittings.
In the following we will use the condense notation $F$ for $ff$
and thus $v^{ff-ff}_l \equiv v^{F-F}_l$.
These  quantities are connected with the Slater 
(Condon-Shortley) parameters \cite{CS}  
$F^l(5f,5f)$ through the following relation: 
\begin{eqnarray}
 v^{F-F}_l=\frac{4\pi}{2l+1} F^l .
\label{3.1} 
\end{eqnarray}
In particular, the Slater parameters $F^2$, $F^4$, and $F^6$
of Amoretti {\it et al}. \cite{Amo} correspond to
$v^{F-F}_2=14.007$ eV, $v^{F-F}_4=7.091$ eV and $v^{F-F}_6=3.168$ eV.
(In order to obtain the exact term splitting quoted in Table \ref{table2}
of Ref.~\onlinecite{Amo}
we had to scale their Slater parameters by a factor 0.9755.)
Alternatively, the quantities $v^{F-F}_l$ can be calculated
by using the radial dependence of the $5f$ 
electrons ${\cal R}_f$, Eq.~(\ref{2.18a}).
We have done such a calculation and then corrected the
parameters by comparing the splittings of the $f^3$
configuration with experimental data for Pr$^{3+}$ and Nd$^{4+}$
(details are given in Appendix B). We arrived at
\begin{eqnarray}
 & & v^{F-F}_2=18.164, \quad v^{F-F}_4=8.578, \quad v^{F-F}_6=3.362 ,
 \nonumber \\
 & & \zeta_f=0.2547 , \quad \mbox{in eV} .
\label{3.2} 
\end{eqnarray}

After calculating the matrix elements of the Coulomb
repulsion and the spin-orbit coupling, we diagonalize
the matrix 
\begin{eqnarray}
  H_{intra}= V^{(3)}+H_{so}
\label{3.4} 
\end{eqnarray}
and obtain the electronic spectrum of $5f^3$.
The 9 lowest and two highest eigenvalues 
are shown in Table \ref{table2}, 
where for comparison we also quote the spectrum
of free Np ion of Ref.~\onlinecite{Amo}.
%
\begin{table} 
\caption{ 
The 9 lowest and 2 highest eigenvalues of $5f^3$,
calculated with $v^{F-F}_l$, Eq.~(\ref{3.2}). 
$g$ stands for the Land\'e factor. $E^A$ refers
to the calculation of Amoretti {\it et al}., Ref.~\onlinecite{Amo}.
Two highest values of $E^A$
marked by an asterisk were reproduced by our calculation
with the parameters of  Ref.~\onlinecite{Amo}.
\label{table2}     } 
\begin{ruledtabular}
 \begin{tabular}{l c c c c c } 
 & term & deg. & $g$ ($\mu_B$) & $E$, meV & $E^A$, meV \\ 
\hline 
1 & $^4I_{9/2}$  & 10 & 0.7546 & 0 & 0 \\
2 & $^4I_{11/2}$ & 12 & 0.9704 & 635.3  & 657.7 \\
3 & $^4I_{13/2}$ & 14 & 1.0993 & 1204.2 & 1256.2 \\
4 & $^4F_{3/2}$  &  4 & 0.6027 & 1244.2 & 948.3 \\
5 & $^2H_{9/2}$  & 10 & 1.0154 & 1617.7 & 1438.3 \\
6 & $^4F_{5/2}$  &  6 & 1.0067 & 1702.9 & 1399.3 \\
7 & $^4I_{15/2}$ & 16 & 1.1797 & 1715.3 & 1762.3 \\
8 & $^4S_{3/2}$  &  4 & 1.6546 & 1861.6 & 1614.9 \\
9 & $^4F_{7/2}$  &  8 & 1.1195 & 1955.3 & 1697.1 \\
... & ... & ... & ... & ... & ... \\
40 & $^2F_{7/2}$ & 8 & 1.1317 & 7796.6 & 6541.1$^*$ \\
41 & $^2F_{5/2}$ & 6 & 0.8589 & 8008.2 & 6631.6$^*$
 \end{tabular} 
\end{ruledtabular}
\end{table} 
Notice, that in comparison with the spectrum \cite{Amo} of 
Amoretti {\it at al}.,
$^4F_{3/2}$ and $^4F_{5/2}$ are higher than $^4I_{13/2}$ and
$^2H_{9/2}$, which is in better agreement with the sequence
of terms for the $4f^3$ configurations of Pr$^{3+}$ and Nd$^{4+}$
 known from experiment. \cite{La_sp}

\subsection{CEF excitations in the disordered phase ($T>25$~K)}

In the disordered phase ($T>25$~K) the electron density
of the $5f^3$ configuration adopts the cubic ($O_h$) site
symmetry. This density modulation is induced by the cubic
crystal electric field experienced by three $5f$ electrons.
In terms of the multipole intersite expansion (\ref{2.8})
it implies that for a given Np site $\vec{n}$ we treat
12 Np neighbors ($\vec{n}'_1=1-12$) and 8 oxygen neighbors
($\vec{n}'_2=1-8$) in spherical
approximation, i.e. $l'=0$ and $S_{\Lambda'}(\vec{n}')$
reduces to $Y_0^0=1/\sqrt{4\pi}$.
The coefficients $c_{\Lambda'}$, Eq.\ (\ref{2.15}), become
simple,
\begin{eqnarray}
 c_0(i_{\alpha'} j_{\alpha'}) = {\frac{1}{\sqrt{4\pi}}}
 \delta(i_{\alpha'},j_{\alpha'}) .  
\label{3.5} 
\end{eqnarray}
Here we write $0$ for $\Lambda' \equiv (l'=0,\,A_{1g})$.
At the central site $\vec{n}$ we expand the CEF in terms
of SAF's $S_{\Lambda_1}(\vec{n})$, 
$\Lambda_1 \equiv (l,A_{1g})$, 
where $A_{1g}$ stands for the unit representation
of the cubic site group $O_h$.
The selection rules for the coefficients 
$c_{\Lambda}(i_{\alpha},j_{\alpha})$
of the $f-f$ transitions imply that there remain only two
nontrivial functions $S_{\Lambda_1}$
with $l=4$ and $l=6$, which correspond to the
cubic harmonics $K_4(\Omega)$ and $K_6(\Omega)$.
The multipole two-center expansion (\ref{2.8}) becomes
\begin{subequations}
\begin{eqnarray}
 & & V(\vec{R}(\vec{n}),\vec{R}'(\vec{n}'))=\frac{1}{\sqrt{4\pi}}
 \sum_{\Lambda_1} v_{\Lambda_1 0}(\vec{n},\vec{n}';\, r,r')\,
 S_{\Lambda_1}(\vec{n}) , \nonumber \\
 & & \label{3.6a}
\end{eqnarray}
where
\begin{eqnarray}
  & & v_{\Lambda_1 0}(\vec{n},\vec{n}';\,r,r')\,
   =    \nonumber \\
  & &  \frac{1}{\sqrt{4\pi}}
   \int \! d\Omega(\vec{n}) \int \! d\Omega'(\vec{n}')\,
      {\frac{  S_{\Lambda_1}(\hat{n}) }
     {|\vec{R}(\vec{n})-\vec{R}'(\vec{n}')|}} .  \quad 
\label{3.6b}
\end{eqnarray}
\end{subequations}
Here $v_{\Lambda_1 0}(\vec{n},\vec{n}';\,r,r')$ has
a same value 
for all 12 Np neighbors ($\vec{n}'_1=1-12$),
and a same value for all 8 oxygen
neighbors ($\vec{n}'_2=1-8$).
As follows from Eq.~(\ref{2.11}) 
$v_{\Lambda_1 0}(\vec{n},\vec{n}';\,r,r')$
{\it is independent of} $r'$. Eq.~(\ref{2.20}) then
can be written in the following form:
\begin{eqnarray}
 v_{\Lambda_1}^{\alpha \alpha}{\,}_0^{\alpha' \alpha'}(\vec{n}-\vec{n}')=
 v_{\Lambda_1}^{\alpha \alpha}{}_0 \cdot Q_{\alpha'},
\label{3.7}
\end{eqnarray}
where
\begin{subequations}
\begin{eqnarray}
  v_{\Lambda_1}^{\alpha \alpha}{}_0 = 
  \int dr\,r^2\, {\cal R}_{\alpha}^2(r) \,
  v_{\Lambda_1\, 0}(\vec{n},\vec{n}';\, r,{\not  r'}) 
\label{3.8a}
\end{eqnarray}
and
\begin{eqnarray}
  Q_{\alpha'}=\int dr'\,{r'}^2\,{\cal R}_{\alpha'}^2(r') . 
\label{3.8b}
\end{eqnarray}
\end{subequations}
Here the integrations are taken over $0<r'<R_{MT}$,
where $R_{MT}$ is the radius of the muffin-tin sphere
of neptunium or oxygen. (The influence of the interstitial
region will be discussed later.)
$Q_{\alpha'}$ refers to an electron at site $\vec{n}'$ which interacts
with one of the three $5f$ electrons at $\vec{n}$.
We then can perform a summation over all electrons at $\vec{n}'$
and include also to this term the interaction with
the nucleus. This results in replacing $Q_{\alpha'}$
by $e Q_{MT}$ in Eq.~(\ref{3.7}), $Q_{MT}$ and $e$ being the total
charge inside the MT sphere and the electron charge
($e=-1$), respectively.
From Eq.~(\ref{2.11}) it follows that
\begin{subequations}
\begin{eqnarray}
  v_{\Lambda_1}^{\alpha \alpha}{}_{0} =  
  v_{\Lambda_1\, 0}(\vec{n},\vec{n}'; R_{MT},{\not  r}')
  \frac{q_l^{\alpha}}{R_{MT}^l} ,
\label{3.9a}
\end{eqnarray}
where $l$ in the index $\Lambda_1$ is 4 or 6, and
\begin{eqnarray}
 q_l^{\alpha}= \int dr'\,{r'}^{\,(l+2)}\, {\cal R}_{\alpha}^2(r')  .
\label{3.9b}
\end{eqnarray}
\end{subequations}

Therefore, the CEF operator for any neptunium site ($r<R_{MT}^{Np}$) 
can be written explicitly as
\begin{subequations}
\begin{eqnarray}
  V_{CF}(\vec{R}(\vec{n}))= \sum_{l=4,6} B_{l}\, 
  S_{(l,A_{1g})}(\hat{n}) \, 
  \left( \frac{r}{R_{MT}^{Np}} \right)^l ,  
\label{3.10a}
\end{eqnarray}
where 
\begin{eqnarray}
  B_{l}=B_{l}^{Np}+B_{l}^{O} ,
\label{3.10b}
\end{eqnarray}
and
\begin{eqnarray}
  & &B_{l}^{Np}=
  {\frac {12}{\sqrt{4 \pi}}}\, Q_{eff}^{Np}\,e \,
  v_{\Lambda_1\, 0}(\vec{n},\vec{n}'_1; R_{MT}^{Np},{\not  r}') ,
\quad \quad \label{3.10c}  \\
  & &B_{l}^{O}=
  {\frac {8}{\sqrt{4 \pi}}}\, Q_{eff}^{O}\,e \,
  v_{\Lambda_1\, 0}(\vec{n},\vec{n}'_2; R_{MT}^{Np},{\not  r}') .
\label{3.10d}
\end{eqnarray}
\end{subequations}
We quote all relevant parameters of CEF in Table \ref{table3}.
%
\begin{table} 
\caption{ 
Calculated parameters of the CEF.
$v_l^{Np}{\,}_{0}^{Np}=
v_{\Lambda_1\, 0}(\vec{n},\vec{n}'_1; R_{MT}^{Np},{\not  r}')$,
$\vec{n}=0$ is the central Np site, $\vec{n}'_1$ is one of 12
Np nearest neighbors.
$v_l^{Np}{\,}_{0}^{O}=
v_{\Lambda_1\, 0}(\vec{n},\vec{n}'_2; R_{MT}^{Np},{\not  r}')$,
$\vec{n}'_2$ is one of 6 oxygen neighbors.
\label{table3}     } 
\begin{ruledtabular}
 \begin{tabular}{l c c c } 
  &  units & $l=4$ & $l=6$ \\ 
\hline 
 $v_l^{Np}{\,}_{0}^O$     & meV  & -816.7  & 209.7 \\
 $v_l^{Np}{\,}_{0}^{Np}$  & meV  & -26.37  & -6.190 \\
 $q_l^f/(R_{MT}^{Np})^l$  &      & 0.1592   & 0.0994 \\
 $B_l^O/Q_{eff}^O\,e$       & K & -3405.2 & 546.0 \\
 $B_l^{Np}/Q_{eff}^{Np}\,e$ & K & -165.0  & -24.2 \\
 \end{tabular} 
\end{ruledtabular}
\end{table} 
As given by Eq.~(\ref{3.10a}), the CEF operator $V_{CF}$ is
a one-electron quantity. \cite{New1,New2} CEF acts along with 
the Coulomb intrasite repulsion, Eq.\ (\ref{3.4}).
Therefore, the total Hamiltonian for the disordered phase
becomes
\begin{eqnarray}
   H^{dis}(\vec{n})= H_{intra} + V_{CF}(\vec{n}) .
\label{3.11}
\end{eqnarray}

Although we have considered CEF from first principles 
there is still an ambiguity related to the charge
distribution in the interstitial region.
A more rigorous treatment of the problem is given in 
Refs.\ \onlinecite{Wei,ND}.
A careful consideration of the problem based on the solution 
of a periodic Poisson's equation leads to 
a renormalization of the charges inside the MT spheres, 
Ref.~\onlinecite{ND}. In other words, in Eqs.\ (\ref{3.10a}-d)
the effective charges for Np and O are given by 
\begin{eqnarray}
  Q_{eff} = 
 Q_{MT}-\frac{4\pi R_{MT}^3}{3} \rho_I(\vec{K}=0) \nonumber \\
     -4\pi R_{MT}^2 {\sum_{\vec{K} \neq 0}}' 
      \frac{j_1(KR_{MT})}{K} \rho_I(\vec{K}) ,
   \label{3.12'}  
\end{eqnarray}
where $\rho_I(\vec{K})$ is the Fourier series expansion of
the electron density in the interstitial region, $j_{l=1}$ is
the spherical Bessel function. $\rho_I(\vec{K}=0)$ is the
average density in the interstitial region, $Q_{out}/V_{out}$,
where $Q_{out}$ and $V_{out}$ is the charge and volume of 
the interstitial region. 

The calculation of $Q_{eff}$ according to Eq.~(\ref{3.12'})
is quite laborious since it requires the evaluation
of the Fourier coefficients $\rho_I(\vec{K})$.
Instead, below we consider two approximation to (\ref{3.12'}). 
In the first approximation 
we assume that $Q_{eff}^{Np}(I)=Q_{MT}^{Np}$ and 
$Q_{eff}^{O}(I)=Q_{MT}^{O}$,
where $Q_{MT}^{Np}$ and $Q_{MT}^{O}$
are the total charges inside the MT spheres of neptunium and oxygen.
Here the electron charge in the interstices is completely ignored.
In the second approximation we take
\begin{eqnarray}
  Q_{eff}(II) = 
 Q_{MT}-\frac{4\pi R_{MT}^3}{3} \rho_I(\vec{K}=0) .
   \label{3.14}  
\end{eqnarray}
This expression corresponds to the homogeneous electron density
distribution in the interstitial region.
However, the modification of effective charges in this approximation 
is too strong: $Q_{eff}^{Np}(I)=+4.108|e|$ and 
$Q_{eff}^{Np}(II)=+5.337|e|$
for neptunium, $Q_{eff}^{O}(I)=-0.284|e|$ and 
$Q_{eff}^{O}(II)=+0.945|e|$ for oxygen.
As was discussed in Refs.~\onlinecite{NM5,NM3}
the CEF splitting is overestimated in the second approximation.
In reality the charge density in interstices is highly
inhomogeneous, concentrated mainly in the proximity to oxygen
and neptunium.
This leads to $\rho_I(\vec{K} \neq 0) \neq 0$, and
the last term in Eq.~(\ref{3.12'}) acts in the opposite
direction, decreasing $Q_{eff}$ backward to $Q_{MT}$ values,
which correspond to the first approximation.

\begin{figure} 
\resizebox{0.46\textwidth}{!}
{ 
 \includegraphics{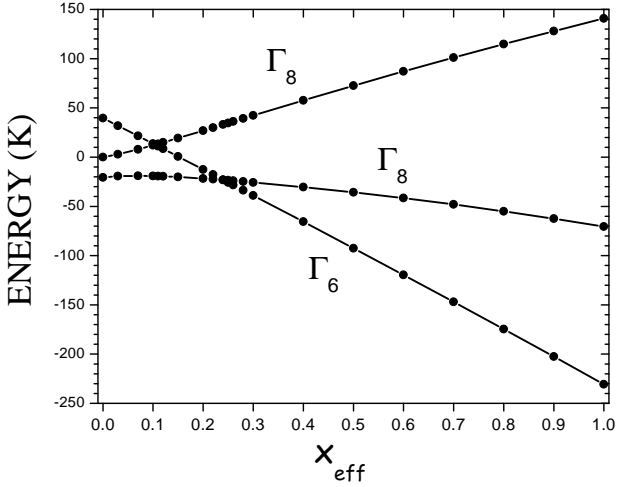} 
} 
\vspace{2mm}
\caption{
Splitting of the lowest $^4 I_{9/2}$ terms of the $5f^3$
configuration of Np in cubic crystal field, Eqs.~(\ref{3.10a}-d)
as a function of the effective charges of neptunium and oxygen,
Eq.~(\ref{3.15a},b). Zero corresponds to the energy of
the $^4 I_{9/2}$ level of free Np ion.
} 
\label{fig2} 
\end{figure}

The exact calculation of $Q_{eff}$ according to Eq.~(\ref{3.12'})
is beyond the scope of the present study. Instead, we
have studied the crystal field effects as a function of $Q_{eff}$
by introducing 
\begin{subequations}
\begin{eqnarray}
  Q_{eff}^{Np}(x_{eff}) = 
   Q_{MT}^{Np}+x_{eff}(Q_{eff}^{Np}(II)-Q_{MT}^{Np}) , \quad \quad
   \label{3.15a}  \\
  Q_{eff}^{O}(x_{eff}) = 
   Q_{MT}^{O}+x_{eff}(Q_{eff}^{O}(II)-Q_{MT}^{O}) , \quad \quad
   \label{3.15b}
\end{eqnarray}
\end{subequations}
where $0<x_{eff}<1$.
Diagonalizing $H^{dis}$, Eq.~(\ref{3.11}), we have found that 41 terms
of $5f^3$ are split into 120
distinct sublevels of $\Gamma_6$, $\Gamma_7$ and $\Gamma_8$
symmetry of the cubic double group $O'_h$. 
In particular two lowest
atomic-like levels are split according to the following scheme,
\begin{subequations}
\begin{eqnarray}
& & ^4 I_{9/2} \rightarrow \Gamma_8+\Gamma_8+\Gamma_6 ,   \label{3.12a} \\
& & ^4 I_{11/2} \rightarrow \Gamma_8+\Gamma_7+\Gamma_6+\Gamma_8 .
               \label{3.12b} 
\end{eqnarray}
\end{subequations}
The resulting splittings and the dependence of CEF on 
$x_{eff}$ is shown in Fig.~\ref{fig2}.
The splittings of two lowest terms of the $5f^3$ configuration,
Eq.~(\ref{3.12a},b), is also given in Tables \ref{table4} and \ref{table5} 
for $x_{eff}=0$
and $x_{eff}=0.5$, respectively.
%
%
%
\begin{table} 
\caption{ 
CEF low energy spectrum and magnetic moments of the $5f^3$
configuration of Np; $x_{eff}=0$, $\triangle \epsilon=7380.7$~K.
Calculated CEF parameters $B_4=-288.1$~K, $B_6=254.2$~K, 
Eq.~(\ref{3.10a}). 
\label{table4}     } 
\begin{ruledtabular}
 \begin{tabular}{l c c c c } 
 & $\Gamma$  & deg. & $(\epsilon_i-\epsilon_1)$ (K) & ${\cal M}_z$($\mu_B$) \\ 
\hline 
 & $\Gamma_8$ & 4 & 0    & $\pm (1.275,\,1.429)$ \\
$^4 I_{9/2}$ & $\Gamma_8$ & 4 & 20.5 & $\pm (0.517,\,1.686)$ \\
 & $\Gamma_6$ & 2 & 60.2 & $\pm 1.384$ \\
\hline
 & $\Gamma_6$ & 2 & $\triangle \epsilon$      & $\pm 1.778$ \\
$^4 I_{11/2}$ & $\Gamma_8$ & 4 & $\triangle \epsilon+4.6$  & $\pm (1.241,\,2.120)$ \\
 & $\Gamma_7$ & 2 & $\triangle \epsilon+6.3$  & $\pm 1.775$ \\
 & $\Gamma_8$ & 4 & $\triangle \epsilon+26.1$ & $\pm (1.119,\,3.835)$
 \end{tabular} 
\end{ruledtabular}
\end{table} 
%
\begin{table} 
\caption{ 
CEF low energy spectrum and magnetic moments of the $5f^3$
configuration of Np; $x_{eff}=0.5$, $\triangle \epsilon=7396.3$~K.
Calculated CEF parameters $B_4=1905.7$~K, $B_6=-66.5$~K, 
Eq.~(\ref{3.10a}).
\label{table5}     } 
\begin{ruledtabular}
 \begin{tabular}{l c c c c } 
 & $\Gamma$  & deg. & $(\epsilon_i-\epsilon_1)$ (K) & ${\cal M}_z$($\mu_B$) \\ 
\hline 
 & $\Gamma_6$ & 2 & 0     & $\pm 1.383$ \\
$^4 I_{9/2}$ & $\Gamma_8$ & 4 & 56.8  & $\pm (0.451,\,2.077)$ \\
 & $\Gamma_8$ & 4 & 164.9 & $\pm (1.204,\,2.333)$ \\
\hline
 & $\Gamma_8$ & 4 & $\triangle \epsilon$       & $\pm (0.392,\,3.862)$ \\
$^4 I_{11/2}$ & $\Gamma_7$ & 2 & $\triangle \epsilon+28.6$  & $\pm 1.777$ \\
 & $\Gamma_8$ & 4 & $\triangle \epsilon+110.8$ & $\pm (1.290,\,2.845)$ \\
 & $\Gamma_6$ & 2 & $\triangle \epsilon+132.6$ & $\pm 1.785$
 \end{tabular} 
\end{ruledtabular}
\end{table} 

The most comprehensive study of the crystal field of the 
$5f^3$ configuration
was performed by Amoretti {\it et al.}, Ref.~\onlinecite{Amo}.
Comparing their results with ours,
we obtain the following relations connecting $B_4$ and $B_6$
with $V_4$ and $V_6$ used there,
\begin{subequations}
\begin{eqnarray}
 B_4 &=& 8\sqrt{\frac{12}{7}} V_4 ,  \label{3.16a}   \\
 B_6 &=& 16\sqrt{8} V_6 .  \label{3.16b}   
\end{eqnarray}
\end{subequations}
We observe that for a realistic choice of $Q_{eff}$,
which corresponds to $x_{eff} \sim 0-0.5$, the CEF splitting is {\it a few times
smaller} than the value 55~meV considered for CEF excitations in 
Ref.~\onlinecite{Amo}. Correspondingly, the calculated parameters
$B_4$ and $B_6$ (Tables \ref{table4} and \ref{table5}) 
are smaller.
We discuss a possible explanation to this fact in the Conclusions,
Sec.~VI. Notice that it is not possible to relate the feature at 
55 meV with the $^4 I_{11/2}$ splittings because it is
situated at much higher energy $\sim 650$ meV, Table \ref{table2}.
Most likely, the experimental excitations at 55 meV
refer to the valence electrons delocalized on the Np-O bonds,
while the lowest CEF excitations of Np lie at a smaller energy $\sim 6$ meV,
Tables \ref{table4}, \ref{table5}. 

Finally, we would like to notice that our CEF calculation 
is based on first principles,
and in that respect it differs from the others, \cite{Amo,Lea,Ste}
which use fitting from experiment to extract the CEF parameters.
There are also other technical differences.
In contrast to Refs.~\onlinecite{Lea,Ste} we do not
assume that the full momentum $J$ is a good quantum number,
that allows for a mixing of components belonging to different $J$.
In Ref.~\onlinecite{Amo} the basis was
truncated to the first 11 low-lying levels.
In our approach we do not have these limitations.

\subsection{Mean field and the structural phase transition at 25 K}

Now we consider the intersite quadrupole interactions 
$V^{QQ}(\vec{n},\vec{n}')$ between 
a central Np site (sublattice $\{ n_1 \}$) and its 12 nearest Np 
neighbors belonging to sublattices $\{n_{p'}\}$, $p'=2,3,4$.
(Here we will not take into account the interactions involving
higher spherical harmonics because they are considerably smaller,
Eq.~(\ref{2.11}).)
We will handle the $Pn{\bar 3}m$ and $Pa{\bar 3}$ symmetries together
since both of them result in the same local quadrupolar function
(\ref{3c.1a}) for neptunium.

The problem for the $Pa{\bar 3}$ structure
has been considered in Refs.~\onlinecite{NM1,NM3}
where it is shown that this spatial order of quadrupoles
gives an effective attraction between them.
This direct quadrupole-quadrupole coupling can be calculated
from first principles. 
However, we will see that its strength
is not sufficient and we have to assume a substantial reinforcement via
oxygen mediated interaction.
For the $Pn{\bar 3}m$ structure the direct interaction is
repulsive and we have to resort to the indirect coupling
from the very beginning. The important fact which we
exploit in this section is that
{\it the rhombohedral (trigonal) mean field for both
structures can be described by the same expression} 
(\ref{3c.7}) below.

We start by deriving an effective mean field for direct Coulomb
coupling between quadrupoles in the $Pa{\bar 3}$ structure.
Following Ref.~\onlinecite{NM3}, one obtains from Eq.~(\ref{2.19})
the following expression for the quadrupolar interaction operator
between neptunium sites at $\vec{n}_1$ and $\vec{n}_{p'}$
\begin{eqnarray}
  V^{Q\, Q}(\vec{n}_1, \vec{n}_{p'}) =       
  -\frac{\gamma^{f \, f} }{3} \, 
   \rho_{f}^Q(\vec{n}_1) \, \rho_{f}^Q(\vec{n}_{p'}) .
\quad \label{3c.2}
\end{eqnarray}
Here $\vec{n}_1 \in \{ n_1 \}$, $\vec{n}_{p'} \in \{ n_{p'} \}$,
and the quadrupolar density operator 
$\rho_{f}^Q$ at site $\vec{n}_p$ is given by
\begin{eqnarray}
  \rho^Q_{f}(\vec{n}) = \sum_{I,J}
   | I \rangle  \,\sum_a P(a) \sum_{\alpha=1}^3 
   c_{p}(i_{\alpha} j^a_{\alpha})\, \prod_{\beta=1}^2 
   \delta_{i_{\beta} j^a_{\beta}} 
   \langle J | , \nonumber \\
\label{3c.3} 
\end{eqnarray}
where again $a$ is a permutation of $j^f_1,j^f_2,j^f_3$
transforming them to a new order given by $j^a_{\kappa}$, $\kappa=1-3$.
$P(a)$ is the parity of the permutation, the index $\alpha$
stands for the interacting electron at the site. 
(The second interacting electron belongs to a neighboring Np site.) 
The other (noninteracting) electrons ($\beta=1,2$) at $\vec{n}$ 
produce the product of the Kronecker delta symbols.
The quadrupolar $f-f$ coefficients are defined as
\begin{eqnarray}
  c_{p}(i_{\alpha} j_{\alpha}) = 
  \langle i_{\alpha} | {\cal S}_{p} | j_{\alpha} \rangle .
\label{3c.4} 
\end{eqnarray} 
There are four types of such coefficients (i.e. $p=1-4$) as follows 
from Eqs.~(\ref{3c.1a}-d).
Finally, the interaction constant $\gamma^{f \, f}$ 
in Eq.~(\ref{3c.2}) is given by
\begin{eqnarray}
  \gamma^{f \, f} =
  \int \! dr\, r^2 \int \! dr'\, {r'}^2\, 
  {\cal R}_f^2(r)\, {\cal R}_f^2(r')\,
  v_{\Lambda \Lambda} (\vec{n},\vec{n}';\, r,r') ,
  \nonumber \\
\label{3c.5} 
\end{eqnarray}
with $v_{\Lambda \Lambda} (\vec{n},\vec{n}';\, r,r')$
defined by Eq.~(\ref{2.9}),
where $\vec{n}=(0,0,0)$, $\vec{n}'=(a/2)(0,1,1)$ and 
$\Lambda=(l=2,\,T_{2g},k=1)$. The corresponding
SAF is $S_{\Lambda}=Y_2^{1s}$.
Using the property (\ref{2.11}) we rewrite $\gamma^{f \, f}$ as
\begin{eqnarray}
  \gamma^{f \, f} = \frac{q_2^f}{(R_{MT}^{Np})^2}\,
  v^{Q\,Q}(R_{MT}^{Np})\, \frac{q_2^f}{(R_{MT}^{Np})^2} ,
\label{3c.6} 
\end{eqnarray}
where the short notation $v^{Q\,Q}(R_{MT}^{Np})$ stands
for $v_{\Lambda \Lambda} (\vec{n},\vec{n}';\, R_{MT}^{Np},R_{MT}^{Np})$,
and the ``quadrupole charge" $q_{l=2}^f$ of $5f$ electron
is given by Eq.~(\ref{3.9b}).

In the mean-field approximation after summing over 12 nearest
neighbors belonging to 3 sublattices we arrive at the
effective bilinear quadrupole-quadrupole operator
\begin{eqnarray}
  U^{Q\, Q}(\vec{n}_p)=-\lambda^{f\, f} \, 
  \langle \rho_f^Q \rangle \, \rho_f^Q(\vec{n}_p) ,
\label{3c.7}
\end{eqnarray}
where $\lambda^{f\, f}=4\gamma^{f\, f}>0$ and the 
calculated parameters are quoted in Table \ref{table6}.
%
\begin{table} 
\caption{ 
Calculated parameters of the mean field.
\label{table6}     } 
\begin{ruledtabular}
 \begin{tabular}{c c c c } 
 $v^{Q\,Q}(R_{MT}^{Np})$ & $q_2^f/(R_{MT}^{Np})^2$ & 
           $\gamma^{f\, f}$ & $\lambda^{f\, f}$ \\
 4568.2 K & 0.3095 & 437.6 K & 1750.5 K
 \end{tabular} 
\end{ruledtabular}
\end{table} 
The same expression holds for $Pn{\bar 3}m$, but in that case
for the direct quadrupole interaction $\lambda^{f\, f}<0$ (repulsion),
Appendix A.
$\langle \rho_f^Q \rangle$
stands for an expectation value of the quadrupole operator (\ref{3c.3}).
At zero temperature it is the quantum average 
\begin{eqnarray}
 \langle \rho_f^Q \rangle =
 \langle I_{gs} | \rho_f^Q(\vec{n}_p) | I_{gs} \rangle ,
\label{3c.8}
\end{eqnarray}
where $| I_{gs} \rangle$ refers to the ground state of
the full mean-field Hamiltonian
\begin{eqnarray}
  H^{MF}(\vec{n})=U^{Q\, Q}(\vec{n}) + 
  V_{CF}(\vec{n}) + H_{intra}(\vec{n}) .
\label{3c.9}
\end{eqnarray}
The intrasite part of the interactions $H_{intra}$ is given
by Eq.~(\ref{3.4}). For CEF we used the values $x_{eff}=0$
and $Q_{eff}^{Np}=Q_{MT}^{Np}$, $Q_{eff}^{O}=Q_{MT}^{O}$.
(As we discussed in Sec.~IV.B this gives the most realistic 
estimate for CEF.)
For both $Pa{\bar 3}$ and $Pm{\bar 3}n$ structures
the mean-field Hamiltonian has the  $S_6=C_3 \times i$ 
trigonal common site symmetry, and lifts the 4-fold
degeneracy of the quartet states,
\begin{eqnarray}
  \Gamma_8 \rightarrow E + E ,
\label{3c.9m}
\end{eqnarray}
where $E$ stands for the two-fold degenerate irreducible
representation of $C_3$. The states $\Gamma_6$ remain
unsplit as a consequence of the Kramers theorem,
$\Gamma_6 \rightarrow E$.

%
\begin{figure} 
\resizebox{0.43\textwidth}{!}
{ 
 \includegraphics{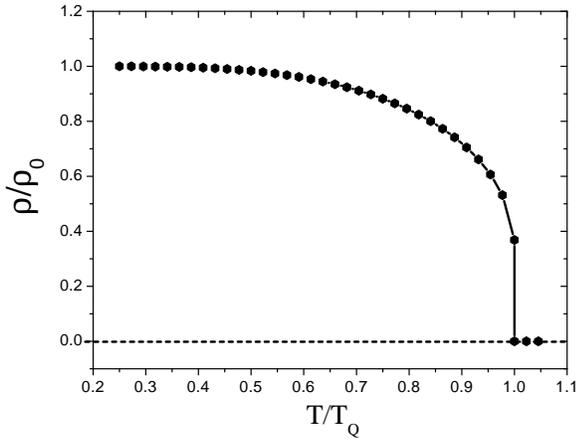} 
} 
\vspace{2mm}
\caption{
A typical evolution of the order parameter amplitude
$\langle \rho_f^Q \rangle$ with temperature;
$\rho_0=\langle \rho_f^Q \rangle |_{T=0}$,
$T_Q$ is the transition temperature.
} 
\label{fig3} 
\end{figure} 
%
We now obtain a system of equations which can
be solved self-consistently.
At first we introduce an average $\langle \rho_f^Q \rangle$,
which defines the interactions (\ref{3c.7}), (\ref{3c.9}).
In the space of the $5f^3$ configuration (364 state vectors),
we diagonalize the total Hamiltonian $H^{MF}$, Eq.~(\ref{3c.9}),
and obtain the eigenvectors $| K \rangle$, $K=1-364$:
\begin{eqnarray}
  H^{MF} | K \rangle = \epsilon_K | K \rangle ,
\label{3c.10}
\end{eqnarray}
where the lowest value of $\epsilon_K$ corresponds to $K=1,2$.
This is the Kramers doublet of the ground state.
We then calculate the quantities 
\begin{eqnarray}
 \rho_f^Q(K)=\langle K | \rho_f^Q(\vec{n}_p) | K \rangle ,
\label{3c.11}
\end{eqnarray}
which evaluate the quadrupolar moments of the states $K$.
Next, we find an improved value for 
$\langle \rho_f^Q \rangle$ which is $\rho_f^Q(K=1)=\rho_f^Q(K=2)$, 
Eq.~(\ref{3c.8}).
The procedure continues until the input and output values
for $\langle \rho_f^Q \rangle$ converge.
As a consequence of symmetry
the expectation value $\langle \rho_f^Q \rangle$
is independent of the sublattice $\{ n_p \}$ chosen
for calculations, but the Hamiltonian and eigenvectors
do depend on the choice. This is because the
quadrupoles have different orientations for different
sublattices.

For nonzero temperature $T$ the mean-field equation for
$\langle \rho_f^Q \rangle$ becomes
\begin{subequations}
\begin{eqnarray}
  \langle \rho_f^Q \rangle = \frac{1}{Z}
    \sum_{K} \rho_f^Q(K) \, e^{-\epsilon_K/T} ,
\label{3c.12a} 
\end{eqnarray}
where
\begin{eqnarray}
   Z=\sum_K e^{-\epsilon_K/T} .
\label{3c.12b} 
\end{eqnarray}
\end{subequations}

%
\begin{table} 
\caption{ 
Mean-field (trigonal) splittings at $T=0$, 
$\lambda_{eff}/\lambda^{f\,f}=1$. 
$T_Q=0.44$~K, $\rho_0=-0.0205$.
\label{table7}     } 
\begin{ruledtabular}
 \begin{tabular}{l c c c r} 
$\Gamma$  & deg. & $(\epsilon_i-\epsilon_1)$ (K) & 
   ${\cal M}_z$($\mu_B$) & $\rho_f^Q$ \\ 
\hline 
 $E$ & 2 & 0    & $\pm0.9464$ & -0.0205 \\
 $E$ & 2 & 1.1  & $\pm0.7962$ &  0.0135 \\
 $E$ & 2 & 20.7 & $\pm0.6406$ & -0.0083 \\
 $E$ & 2 & 21.7 & $\pm0.8428$ &  0.0140 \\
 $E$ & 2 & 60.8 & $\pm1.3840$ &  0.0011
 \end{tabular} 
\end{ruledtabular}
\end{table} 
\begin{figure} 
\resizebox{0.40\textwidth}{!}
{ 
 \includegraphics{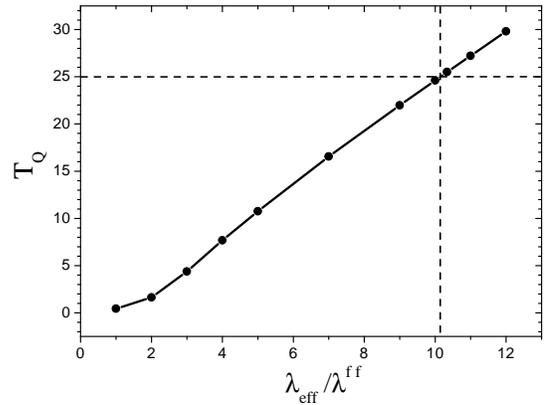} 
} 
\vspace{2mm}
\caption{
The transition temperature $T_Q$ as a function of
the quadrupolar interaction constant $\lambda_{eff}$.
} 
\label{fig4} 
\end{figure} 
%
%
\begin{table} 
\caption{ 
Mean-field (trigonal) splittings at $T=0$ with 
an enhanced quadrupolar interaction constant 
$\lambda_{eff}/\lambda^{f\,f}=10$. 
$T_Q=24.6$~K, $\rho_0=-0.0563$.
\label{table8}     } 
\begin{ruledtabular}
 \begin{tabular}{l c c c r} 
$\Gamma$  & deg. & $(\epsilon_i-\epsilon_1)$ (K) & 
   ${\cal M}_z$($\mu_B$) & $\rho_f^Q$ \\ 
\hline 
 $E$ & 2 & 0     & $\pm1.9211$ & -0.0563 \\
 $E$ & 2 & 54.0  & $\pm1.1723$ & -0.0014 \\
 $E$ & 2 & 72.8  & $\pm1.3684$ & -0.0004 \\
 $E$ & 2 & 85.8  & $\pm0.6211$ &  0.0249 \\
 $E$ & 2 & 122.1 & $\pm1.4893$ &  0.0256
 \end{tabular} 
\end{ruledtabular}
\end{table} 
The results of the calculations are quoted in
Tables \ref{table7}, \ref{table8}, and in Figs.~3, 4, 5.
With $\lambda^{f\, f}=$1750.5~K, Table \ref{table6}, 
which corresponds to the direct quadrupole-quadrupole attraction 
of the $Pa{\bar 3}$ structure,
we found that the transition temperature $T_Q$
is only 0.44~K.
A typical dependence of $\langle \rho_f^Q \rangle$ is shown
in Fig.~3. The phase transition is of first order,
with a discontinuity of the order parameter
amplitude $\langle \rho_f^Q \rangle=-0.0075$ at $T_Q$.
Comparing the present calculation with that for cerium, \cite{NM3}
we observe that the low value of $T_Q$ is due to a
small quadrupolar susceptibility of the ground state 
quartet $\Gamma_8$, since the difference in $\lambda$ is
not that much (for cerium $\lambda^{f\, f}=2241$~K.)

The calculated transition temperature is very small
in comparison with the experimental value of 25~K.
We then conclude that in the framework of the 
$5f^3$ model
the structural phase transition
can not be explained by means of the direct bilinear quadrupole
coupling.
Most likely, there is indirect coupling between $5f$
electron densities on neptunium
via the oxygen atoms (superexchange interaction \cite{super}).
The microscopic consideration of the superexchange interaction
is beyond the scope of the present work.
Instead, we model it by increasing the value of $\lambda^{f\, f}$.
In such a case $\lambda^{f\, f}$ becomes
a phenomenological parameter which we denote as $\lambda_{eff}$.
By changing $\lambda_{eff}$ we change the transition temperature
as shown in Fig.~4.
We have found that the experimental value of 25 K is achieved
for $\lambda_{eff}/\lambda^{f\,f} \sim 10$, which indicates a substantial
increase of the effective bilinear coupling, Eq.~(\ref{3c.7}).
The relevant parameters of such strong mean-field are
given in Table \ref{table8}.

\begin{figure} 
\resizebox{0.40\textwidth}{!}
{ 
 \includegraphics{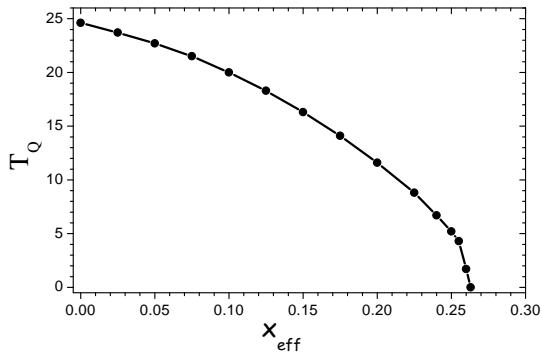} 
} 
\vspace{2mm}
\caption{
The suppression of the transition temperature $T_Q$ with
the increase of the CEF strength, $x_{eff}$, see
also Fig.~2 and Sec.~IV.B; $\lambda_{eff}/\lambda^{f\, f}=10$.
} 
\label{fig5} 
\end{figure}

Finally, we would like to mention that the mean-field
calculations have been done assuming that
the CEF is weak, i.e. $x_{eff}=0$.
Increasing $x_{eff}$ leads to an increase of
the CEF splittings, which results in a strong suppression
of the transition temperature $T_Q$, Fig.~\ref{fig5}.
Notice that at $x_{eff}(\Gamma_8 \rightarrow \Gamma_6)=0.241$ 
the ground state changes
to the $\Gamma_6$ doublet, Fig.~\ref{fig2}, which is
apparently unfavorable for the quadrupolar order. With further
increase of $x_{eff}$ beyond the $x_{eff}(\Gamma_8 \rightarrow \Gamma_6)$
point the transition temperature goes fast to zero
and the ordered phase disappears.
%

\section {Four electron configurations at
neptunium site} 

As follows from the previous section
the model which takes into account only the $5f^3$ 
configuration at each neptunium site is not capable
to explain the disappearance of the magnetic moments
in the ordered phase.
On the other hand, the charge distribution inside
the MT sphere centered at the neptunium nucleus indicates that 
there is always approximately one 
valence electron instantaneously
present along with the three localized $5f$ electrons,
Table~\ref{table1}.
Even if the electron is in the $s$-orbital state, it 
experiences strong coupling with the $5f$ electrons via
intrasite $s-f$ transitions.
Therefore, the excitation spectrum at each neptunium site
differs from that for $5f^3$ considered in Sec.~IV. 
In the following we model
the couplings with the valence electron by considering
$7s5f^3$, $7p5f^3$, and $6d5f^3$ instantaneous configurations.
Here we will not study the mean-field in such detail as
for the $5f^3$ configuration. Our main objective is to
show that the ground state can be nonmagnetic and separated
from the magnetic excitations by an energy gap larger
than 25~K.

\subsection{$7s5f^3$ configuration}

The $7s5f^3$ configuration has been considered in detail in
Sec.~III. We are working here in the space of 728 state vectors
$|I \rangle = | i_1^f, i_2^f, i_3^f;\, i^s \rangle$. 
The $5f$-states are coupled to two $s-$ states through
the $f-s$ transitions accompanied by the multipole
Coulomb interactions with $l=3$. The strength of this interaction
was estimated from the LDA calculation of a Np atom, Eq.~(\ref{2.18b}),
\begin{eqnarray}
 & & v^{fs-sf}_3=1.313  
 \quad \mbox{eV} . \label{4a.1} 
\end{eqnarray}

The electron energy spectrum of the $7s5f^3$ configuration consists
of 82 distinct levels, see Table \ref{table9}.
%
\begin{table} 
\caption{ 
The 5 lowest and the highest eigenvalue of $7s5f^3$;
$g$ is the Land\'e factor. 
\label{table9}     } 
\begin{ruledtabular}
 \begin{tabular}{l c c c r } 
 & term & deg. & $g$ ($\mu_B$) & $E$ \\ 
\hline 
1 & $^5I_4$  &  9 & 0.6266 & 0 \\
2 & $^5I_5$  & 11 & 0.8983 & 196.5  \\
3 & $^5I_6$  & 13 & 1.0670 & 737.2 \\
4 & $^5F_5$  & 11 & 0.8700 & 814.4 \\
5 & $^5I_7$  & 15 & 1.1691 & 1245.1 \\
... & ...   & ... & ...    & ...     \\
82 & $^1F_0$ & 7  & 1.0172 & 8229.5
 \end{tabular} 
\end{ruledtabular}
\end{table} 
In the cubic crystal field two lowest levels ($^5I_4$ and $^5I_5$)
are split as quoted in Table \ref{table10}.
%
\begin{table} 
\caption{ 
CEF low energy spectrum and magnetic moments of the $7s5f^3$
configuration of Np, $\triangle \epsilon=2293.2$~K.
CEF parameters $B_4=-288.1$~K, $B_6=254.2$~K; $x_{eff}=0$.
\label{table10}     } 
\begin{ruledtabular}
 \begin{tabular}{l l c c c } 
 & $\Gamma$  & deg. & $(\epsilon_i-\epsilon_1)$ (K) & ${\cal M}_z$($\mu_B$) \\ 
\hline 
 & $E$   & 2 & 0    & 0; 0 \\
$^5 I_4$ & $T_2$ & 3 & 15.6 & $\pm1.5671$; 0 \\
 & $T_1$ & 3 & 25.1 & $\pm0.2946$; 0 \\
 & $A$   & 1 & 61.6 &  0 \\
\hline
 & $T_1$ & 3 & $\triangle \epsilon$      &  $\pm2.2435$; 0 \\
$^5 I_5$ & $E$   & 2 & $\triangle \epsilon$+4.1  &  0; 0 \\
 & $T_2$ & 3 & $\triangle \epsilon$+6.5  &  $\pm2.2393$; 0 \\
 & $T_1$ & 3 & $\triangle \epsilon$+20.3 &  $\pm2.7113$; 0 
 \end{tabular} 
\end{ruledtabular}
\end{table} 
Notice that the ground state is nonmagnetic.
However, at higher temperatures ($T>25$~K) two low
lying excitations of $T_2$ and $T_1$ symmetry 
contribute to the Curie law for the magnetic susceptibility. \cite{Erd2}

In the ordered phase the local symmetry of the Np site
changes to $S_6$ ($Pa{\bar 3}$) or $D_{3d}$ ($Pn{\bar 3}m$). 
As we discussed in Sec.~IV.C
the mean-field can be expanded in a multipole series
and from symmetry it follows that the strongest
interaction is of quadrupolar type,
\begin{eqnarray}
  U^{Q\, Q}(\vec{n}_p)=-\Lambda^f \, \rho_f^Q(\vec{n}_p) ,
\label{4a.2}
\end{eqnarray}
where $\Lambda^f=\lambda^{f\, f} \langle \rho_f^Q \rangle$
[compare with Eq.~(\ref{3c.7})].
The quadrupolar operator $\rho_f^Q(\vec{n}_p)$ belonging to
the sublattice $\{ n_p \}$ is given again by Eq.~(\ref{3c.3})
with the corresponding orientational function ${\cal S}_p$,
Eqs (\ref{3c.1a}-d). The
only difference is that the $s-$electron produces an additional
Kronecker symbol, i.e. in  Eq.~(\ref{3c.3}) $\beta=1-3$.
Notice that both CEF, Eq.~(\ref{3.10a}), 
and the mean field, Eq.~(\ref{4a.2}), act only on the $5f$ electrons.
Taking $\Lambda^f=5252$~K we diagonalized the
full Hamiltonian
\begin{eqnarray}
  H^{MF}(\vec{n})=U^{Q\, Q}(\vec{n}) + 
  V_{CF}(\vec{n}) + H_{intra}(\vec{n}) .
\label{4a.3}
\end{eqnarray}
The resultant electronic spectrum is shown in Table \ref{table11}.
%
\begin{table} 
\caption{ 
Mean-field (trigonal) splittings of $7s5f^3$ at $T=0$,
$\Lambda^{f}=5252$~K.
\label{table11}     } 
\begin{ruledtabular}
 \begin{tabular}{l c c c } 
$\Gamma$  & deg. & $(\epsilon_i-\epsilon_1)$ (K) & ${\cal M}_z$($\mu_B$) \\ 
\hline 
 $A$   & 1 & 0     & 0 \\
 $E$   & 2 & 26.0  & $\pm0.3467$ \\
 $E$   & 2 & 124.1 & $\pm0.6786$ \\
 $A$   & 1 & 301.8 &  0 \\
 $A$   & 1 & 330.1 &  0 \\
 $E$   & 2 & 516.5 & $\pm1.4769$ 
 \end{tabular} 
\end{ruledtabular}
\end{table} 
We observe that the ground state is nonmagnetic, while
the first magnetic excitation ($E$) does not contribute
to the magnetic susceptibility if $T<25$~K.

\subsection{$7p5f^3$ configuration}

In case of the $7p5f^3$ configuration, we
construct $(14 \times 13 \times 12/3!) \times 6$=2184 basis vectors
\begin{eqnarray}
 |I \rangle = | i_1^f, i_2^f, i_3^f;\, i^p \rangle .
\label{4b.0} 
\end{eqnarray}
As before, $i^f$ refers to a $5f$ electron,
$i^p=(k,s_z)$ to the $7p$
electron ($k$ is its orbital part, $k=1-3$, and $s_z$
is the spin part).

The interaction between $5f$ electrons was described in
detail earlier. In addition, $p-p$ and $f-f$ transitions lead to
intrasite multipole interactions with the $l=0$ and $l=2$
components, $f-p$ transitions give
the Coulomb multipole couplings with $l=2$ and $l=4$.
The relevant Slater parameters were extracted from the radial
dependences of a Np ion,
\begin{eqnarray}
 & & v^{fp}_2=1.442 , \quad v^{fp}_4=0.739 , \nonumber \\
 & & v^{ff-pp}_2=5.237 \quad \mbox{(in eV)} .
\label{4b.1} 
\end{eqnarray}
Also, the $p$-electron
experiences the spin-orbit interaction with $\zeta_p=1.2795$~eV.

First, we calculated the electron spectrum of the free ion,
$H_{intra}$ and obtained 242 distinct levels, Table \ref{table12}.
%
\begin{table} 
\caption{ 
The 5 lowest and the highest eigenvalue of $7p5f^3$;
$g$ is the Land\'e factor. 
\label{table12}     } 
\begin{ruledtabular}
 \begin{tabular}{l c c c r } 
 & term & deg. & $g$ ($\mu_B$) & $E$ \\ 
\hline 
1 & $^5K_5$  & 11 & 0.7430 & 0 \\
2 & $^5I_4$  &  9 & 0.7459 & 154.5 \\
3 & $^5K_6$  & 13 & 0.9450 & 689.7  \\
4 & $^5I_5$  & 11 & 0.9875 & 756.2  \\
5 & $^5G_2$  &  5 & 0.5948 & 1248.1 \\
... & ...   & ... & ...    & ...    \\
242 &$^3P_1$ & 3  & 0.5031 & 10515.7
 \end{tabular} 
\end{ruledtabular}
\end{table} 
In the cubic crystal the $^5 K_4$ and $^5 I_4$ levels are 
split as quoted
in Table \ref{table12'}.
(The procedure of treating CEF effects is outlined in Sec.~IV.B,
CEF does not act on the $p-$electron.)
%
\begin{table} 
\caption{ 
CEF low energy spectrum and magnetic moments of the $7p5f^3$
configuration of Np, $\triangle \epsilon=1789.1$~K.
CEF parameters $B_4=-288.1$~K, $B_6=254.2$~K; $x_{eff}=0$.
\label{table12'}     } 
\begin{ruledtabular}
 \begin{tabular}{l l c c c } 
 & $\Gamma$  & deg. & $(\epsilon_i-\epsilon_1)$ (K) & ${\cal M}_z$($\mu_B$) \\ 
\hline 
 & $T_2$ & 3 & 0    & $\pm1.8525$; 0 \\
 $^5 K_5$ & $T_1$ & 3 & 3.2  & $\pm1.7788$; 0 \\
 & $E$   & 2 & 16.7 & 0; 0 \\
 & $T_1$ & 3 & 45.0 & $\pm2.1534$; 0 \\
\hline
 & $E$   & 2 & $\triangle \epsilon$      &  0; 0 \\
 $^5 I_4$ & $T_2$ & 3 & $\triangle \epsilon$+14.8 &  $\pm1.8645$; 0 \\
 & $T_1$ & 3 & $\triangle \epsilon$+24.8 &  $\pm0.3655$; 0 \\
 & $A$   & 1 & $\triangle \epsilon$+59.1 &  0 
 \end{tabular} 
\end{ruledtabular}
\end{table} 
Notice, that the ground state now is
a magnetic triplet of the $T_2$ symmetry, which 
together with two other magnetic $T_1$ excitations, Table \ref{table12'},
gives the Curie law for the magnetic susceptibility. 

At $T_c=25$~K the structural phase transition occurs and the
symmetry of the Np sites is reduced.
This symmetry change is accompanied by lifting degeneracies
of some cubic levels. In particular, the ground state $T_2$ triplet
is split in a doublet and a single level as demonstrated
in Table~\ref{table14}. The mean-field is 
approximated by its quadrupolar electric part,
\begin{eqnarray}
  U^{Q\, Q}(\vec{n}_i)=-\Lambda^f \, \rho_f^Q(\vec{n}_i) 
  -\Lambda^p \, \rho_p^Q(\vec{n}_i) ,
\label{4b.2}
\end{eqnarray}
where 
\begin{subequations}
\begin{eqnarray} 
  \Lambda^f = \lambda^{f\, f} \langle \rho_f^Q \rangle 
  +\lambda^{f\, p} \langle \rho_p^Q \rangle , \label{4b.3a} \\
  \Lambda^p = \lambda^{p\, f} \langle \rho_f^Q \rangle 
  +\lambda^{p\, p} \langle \rho_p^Q \rangle .  \label{4b.3b} 
\end{eqnarray}
\end{subequations}
The following parameters of the interaction were assumed,
\begin{eqnarray}
 & &  \frac{q_2^p}{(R_{MT}^{Np})^2}=0.1604 ,
 \quad \Lambda^f=6612 \; \mbox{K},
 \quad \Lambda^p=3426 \; \mbox{K} . \nonumber \\
\label{4b.4} 
\end{eqnarray}
%
\begin{table} 
\caption{ 
Mean-field (trigonal) splittings of $7p5f^3$ at $T=0$;
$\Lambda^{f}=6612$~K, $\Lambda^{p}=3426$~K.
\label{table14}     } 
\begin{ruledtabular}
 \begin{tabular}{l c c c } 
$\Gamma$  & deg. & $(\epsilon_i-\epsilon_1)$ (K) & ${\cal M}_z$($\mu_B$) \\ 
\hline 
 $A$   & 1 & 0     & 0 \\
 $E$   & 2 & 24.6  & $\pm0.4171$ \\
 $E$   & 2 & 105.8 & $\pm0.8134$ \\
 $A$   & 1 & 246.4 &  0 \\
 $A$   & 1 & 272.6 &  0 \\
 $E$   & 2 & 480.7 & $\pm1.7473$ 
 \end{tabular} 
\end{ruledtabular}
\end{table} 
Since the charge density expansion of the $p$-electron
has quadrupolar components, 
$c_k(i^p j^p)=\langle i^p | {\cal S}_k | j^p \rangle \neq 0$
(${\cal S}_k$ are given by Eqs (\ref{3c.1a}-d)), 
the mean-field expansion (\ref{4b.2}) includes the 
quadrupolar projection on $p$-states,
\begin{eqnarray}
  \rho^Q_{p}(\vec{n}_k) = \sum_{I,J}
   | I \rangle  \,\sum_a P(a) 
   c_k(i^p j^p)\, \prod_{\beta=1}^3 
   \delta_{i_{\beta}^f j^a {}_{\beta}^f} 
   \langle J | . \nonumber \\
\label{4b.5} 
\end{eqnarray}
Here $a$ is a permutation of $j^f_1,j^f_2,j^f_3$ to 
$j^a{}^f_{\beta}$, $\beta=1-3$;
$P(a)$ is the parity of the permutation.
The quadrupolar operator for the $5f$ electrons is given
again by Eq.~(\ref{3c.3}) where the index $\beta$ comprises
the additional $p-$electron, i.e. $\beta=1-3$.

Notice that in the ordered phase the ground state level
is single and nonmagnetic, Table \ref{table14}. 
This mechanism can explain
the loss of magnetic moments because the magnetic
excitations of the $7p5f^3$ configuration lie too high
in energy.

\subsection{$6d5f^3$ configuration}

The basis vectors here are
\begin{eqnarray}
  |I \rangle =| i_1^f, i_2^f, i_3^f;\, i^d \rangle ,
\label{4c.0} 
\end{eqnarray}
where index $i^f$ stands for $5f$ states ($i^f=1-14$),
while the index $i^d=(k,s_z)$ refers to five $d$-orbitals
and the spin projection $s_z$.
Thus, $i^d=1-10$, and in total there are 3640 nonequivalent
basis vectors $|I \rangle$.

We start by considering the intrasite interactions $H_{intra}$.
Here, in addition to $f-f$ interactions we distinguish
two groups. The first group arises between $d-d$ and $f-f$ transitions.
It is described by the multipole Coulomb repulsion with the
$l=0$, 2 and 4 angular components (SAF's). 
The second group is due to the $f-d$ and
$d-f$ transitions. The corresponding multipole
interactions are with $l=1$, 3 and 5.
The relevant parameters were extracted from the LDA
calculation of the Np ion in the $6d5f^3$ configuration,
\begin{eqnarray}
 & & v^{fd-df}_1=11.322 , 
 \quad v^{fd-df}_3=3.701 ,
 \quad v^{fd-df}_5=1.794 ,
 \nonumber \\
 & & v^{ff-dd}_2=11.289 ,
 \quad v^{ff-dd}_4=3.482 \quad \mbox{(in eV)}, \nonumber \\
 & & \zeta_d=0.3497 \; \mbox{eV} . 
\label{4c.1} 
\end{eqnarray}
The parameters for the $f-f$ interactions were kept
unchanged.
We then diagonalized the $3640 \times 3640$ matrix of
$\langle I |H_{intra}|J \rangle$ and obtained 383 distinct
levels. The 5 lowest and the highest levels are quoted in
Table \ref{table15}.
%
\begin{table} 
\caption{ 
The 5 lowest and the highest eigenvalue of $6d5f^3$;
$g$ is the Land\'e factor. 
\label{table15}     } 
\begin{ruledtabular}
 \begin{tabular}{l c c c r } 
 & term & deg. & $g$ ($\mu_B$) & $E$ \\ 
\hline 
1 & $^5L_6$  & 13 & 0.7530 & 0 \\
2 & $^5K_5$  & 11 & 0.7201 & 167.2 \\
3 & $^5L_7$  & 15 & 0.9232 & 667.9  \\
4 & $^5K_6$  & 13 & 0.9269 & 767.2  \\
5 & $^3D_3$  &  7 & 0.6800 & 886.9 \\
... & ...    & ...& ...    & ...    \\
383 &$^3P_1$ & 3  & 1.0004 & 13116.4
 \end{tabular} 
\end{ruledtabular}
\end{table} 
The CEF splittings of the lowest $^5L_6$ and $^5K_5$ levels are given
in Table \ref{table16}.
%
\begin{table} 
\caption{ 
CEF low energy spectrum and magnetic moments of the $6d5f^3$
configuration of Np, $\triangle \epsilon=1949.5$~K.
CEF parameters $B_4=-288.1$~K, $B_6=254.2$~K,
and $B_4^{d}=-232.7$~K; $x_{eff}=0$.
\label{table16}     } 
\begin{ruledtabular}
 \begin{tabular}{l l c c c } 
 & $\Gamma$  & deg. & $(\epsilon_i-\epsilon_1)$ (K) & ${\cal M}_z$($\mu_B$) \\ 
\hline 
 & $A$   & 1 & 0    & 0 \\
 & $T_2$ & 3 & 9.7  & $\pm0.1809$; 0 \\
$^5 L_6$ & $A$   & 1 & 12.6 & 0 \\
 & $T_1$ & 3 & 15.6 & $\pm0.3772$; 0 \\
 & $E$   & 2 & 36.4 & 0; 0 \\
 & $T_2$ & 3 & 43.6 & $\pm2.0637$; 0 \\
\hline
 & $T_1$ & 3 & $\triangle \epsilon$      &  $\pm1.7253$; 0 \\
$^5 K_5$ & $E$   & 2 & $\triangle \epsilon$+3.1  &  0; 0 \\
 & $T_2$ & 3 & $\triangle \epsilon$+21.1 &  $\pm1.8006$; 0 \\
 & $T_1$ & 3 & $\triangle \epsilon$+25.3 &  $\pm2.0883$; 0
 \end{tabular} 
\end{ruledtabular}
\end{table} 
It should be noted that unlike before, the CEF operator 
acts not only on the $5f$ electrons but also 
on the $6d$ one, \cite{NM3}
\begin{eqnarray}
  U^{CEF}(\vec{n})=B_4^f \, \rho_f^4(\vec{n})+
  B_6^f \, \rho_f^6(\vec{n}) +
  B_4^d \, \rho_d^4(\vec{n}) . \quad
\label{4c.1'}
\end{eqnarray}
Here $\rho_f^l(\vec{n})$, $l=4,6$, and $\rho_d^4(\vec{n})$
are cubic projectors on the $f$ and $d$ states, respectively,
given by 
\begin{subequations}
\begin{eqnarray}
  \rho^l_{f}(\vec{n}) = \sum_{I,J}
   | I \rangle  \,\sum_a P(a) \sum_{\alpha=1}^3
   \langle i^f_{\alpha} |K_l| j^a{}^f_{\alpha} \rangle \, 
   \prod_{\beta=1}^3 
   \delta_{i_{\beta} j^a_{\beta}} 
   \langle J | , \nonumber \\
\label{4c.1new} \\
  \rho^4_{d}(\vec{n}) = \sum_{I,J}
   | I \rangle  \,\sum_a P(a) 
   \langle i^d |K_4| j^a{}^d \rangle\, 
   \prod_{\beta=1}^3 
   \delta_{i_{\beta}^f j^a{}_{\beta}^f} 
   \langle J | , \nonumber \\
\label{4c.2new} 
\end{eqnarray}
\end{subequations}
where $K_l(\Omega)$ refers to the cubic harmonics with $l=4$ and 6.
Here we keep the same notations as before [Eqs (\ref{3c.3}) and 
(\ref{4b.5})], i.e. the permutation $a$ transforms the indices
$j^f_1,j^f_2,j^f_3$ to $j^a {}^f_{\kappa}$, $\kappa=1-3$.
The permutations which interchange the $d$ and $f$ indices are
excluded because they give zero contribution to (\ref{4c.1new},b).
The parameter $B_4^d$ was calculated by the method described 
in Sec.~IV.B. For $x_{eff}=0$ we have found that
\begin{eqnarray}
 & &  \frac{q_4^d}{(R_{MT}^{Np})^4}=0.1287 ,
 \quad B_4^d=-232.7 \; \mbox{K} . \quad
\label{4c.2} 
\end{eqnarray}
Notice that for the ground state 
the CEF gives a nonmagnetic single level, Table \ref{table16},
but there are two low lying magnetic levels ($T_2$ and $T_1$)
at 9.7 and 15.6 K, which contribute to the Curie law of the
magnetic susceptibility at $T>25$ K.

Below 25~K the local symmetry of Np is lowered. 
The mean field is given by
\begin{eqnarray}
  U^{Q\, Q}(\vec{n}_p)=-\Lambda^f \, \rho_f^Q(\vec{n}_p) 
  -\Lambda^d \, \rho_d^Q(\vec{n}_p) ,
\label{4c.3}
\end{eqnarray}
where 
\begin{subequations}
\begin{eqnarray} 
  & & \Lambda^f = \lambda^{f\, f} \langle \rho_f^Q \rangle 
  +\lambda^{f\, d} \langle \rho_d^Q \rangle , \label{4c.4a} \\
  & & \Lambda^d = \lambda^{d\, f} \langle \rho_f^Q \rangle 
  +\lambda^{d\, d} \langle \rho_d^Q \rangle . \label{4c.4b} 
\end{eqnarray}
\end{subequations}
Here again $\rho_f^Q(\vec{n})$ and $\rho_d^Q(\vec{n})$
are quadrupolar projection on the $f$ and $d$ states,
respectively. They are given by expressions similar
to (\ref{4c.1new}-b), where we replace 
$K_l(\hat{n}_p)$ by ${\cal S}_p$, Eqs (\ref{3c.1a}-d),
for 4 sublattices $\{ n_p \}=1-4$ of $Pa{\bar 3}$ or $Pn{\bar 3}m$,
Sec.~II. 
[Compare also with Eqs (\ref{3c.3}) and (\ref{4b.5}).]
Below we approximated the parameters
of this interaction by
\begin{eqnarray}
 \frac{q_2^d}{(R_{MT}^{Np})^2}=0.1713 ,
 \quad \Lambda^f=6220  \; \mbox{K},
 \quad \Lambda^d=3442  \; \mbox{K} . \nonumber \\
\label{4c.5} 
\end{eqnarray}
We then diagonalized the whole Hamiltonian $H^{MF}$
($H^{MF}=U^{Q\, Q}(\vec{n}) +  V_{CF} + H_{intra}$) 
and obtained the lowest energy levels quoted in Table~\ref{table17}.
%
\begin{table} 
\caption{ 
Mean-field (trigonal) splittings of $6d5f^3$ at $T=0$;
$\Lambda^{f}=6220$~K, $\Lambda^{d}=3442$~K.
\label{table17}     } 
\begin{ruledtabular}
 \begin{tabular}{l c c c } 
$\Gamma$  & deg. & $(\epsilon_i-\epsilon_1)$ (K) & ${\cal M}_z$($\mu_B$) \\ 
\hline 
 $A$   & 1 & 0     & 0 \\
 $E$   & 2 & 32.8  & $\pm0.4362$ \\
 $E$   & 2 & 136.6 & $\pm0.8797$ \\
 $A$   & 1 & 311.2 &  0 \\
 $A$   & 1 & 337.0 &  0 \\
 $E$   & 2 & 601.2 & $\pm1.7673$ 
 \end{tabular} 
\end{ruledtabular}
\end{table} 
Notice that now the first magnetic excitation
of $E$ symmetry is separated from the nonmagnetic ground
state by an energy gap of $\sim33$~K, which implies
again disappearance of the Curie law for the magnetic 
susceptibility of the ordered phase ($T<25$~K).

\section {Discussion and conclusions} 
\label{sec:dc} 

In resonant X-ray scattering (RXS) experiments the transition to the ordered 
phase at $T_c=25.5$~K
in NpO$_2$ manifests itself by the appearance of superstructure Bragg
reflections [like (003) and others], which are not compatible
with the $Fm{\bar 3}m$ structure of the disordered phase. \cite{Man,Pai}
Paix\~{a}o {\it et al.}, Ref.~\onlinecite{Pai}, ascribed the symmetry of
the ordered phase to the $Pn{\bar 3}m$ space group 
(No.~224, Ref.~\onlinecite{Tables}), but did not elaborate
on the theory.
In this work (Sec.~II), besides $Pn{\bar 3}m$ we discuss also 
the space group $Pa{\bar 3}$ (No.~205), 
which is the other possible candidate for the 
triple-$\vec{q}^X$ quadrupole ordering.
$Pn{\bar 3}m$ and $Pa{\bar 3}$ are close symmetries. They imply
condensations of
different modes ($X_4^+$ and $X_5^+$, Ref.~\onlinecite{Sto})
at the same $X-$point of the Brillouin zone.
We have considered the dependence of the
scattering amplitude for different polarizations on the azimuthal angle
$\psi$ and the Bragg angle $\Theta$
taking into account the domain pattern of both symmetries.
$Pn{\bar 3}m$ and $Pa{\bar 3}$ produce superstructure 
Bragg reflections
at the same sites $(h,k,l)$ of the reciprocal lattice
(Tables \ref{table0a} and \ref{table0b}).
The relation between the space groups is such that in
RXS experiments the $(00 \ell)$ reflection
of $Pn{\bar 3}m$ behaves like the $(\ell 00)$ and $(0 \ell 0)$ 
reflections of $Pa{\bar 3}$ ($\ell=2n+1$).
Therefore, we believe that a special
care should be taken to distinguish between these two symmetries.
We have shown that the two structures are completely different 
with respect to the direct bilinear quadrupole-quadrupole interactions.
The $Pn{\bar 3}m$ symmetry leads
to a repulsion between the quadrupoles while
the $Pa{\bar 3}$ structure implies an attraction.
(This is the main reason why the $Pa{\bar 3}$ symmetry occurs
in many molecular solids: NaO$_2$ \cite{Zie}, N$_2$ \cite{Scott},
C$_{60}$. \cite{Hei2,Sach1,Dav,Sach2,Mic1})

We have presented an {\it ab initio} approach to crystal- and
mean field for the structural phase transition at 25~K.
The method is based on the technique of expanding the Coulomb
repulsion between electrons in a multipolar series, Sec.~III. 

In the disordered phase we considered the Hamiltonian
which includes the crystal electric field (CEF) effects
and the intrasite Coulomb repulsion responsible for
Hund's rules on equal footing, Sec.~IV.
The crystal electric field levels above the ground state 
are in fact the lowest excitations of the neptunium electron complex. 
The typical splittings there are of the order of ten kelvins 
and thus
the electron spectrum is very sensitive to the crystal symmetry.
In this paper we have pursued the simple idea that the
symmetry lowering produces a splitting of the many electron spectrum,
which can explain the difference in the behavior of the magnetic
susceptibility and the loss of the magnetic moments.
Our band structure calculation (Sec.~III.A)
indicates that besides the three localized $5f$ electrons
there is always approximately one conduction electron at 
each Np site.
This changes the effective
instantaneous configuration from the three electron $5f^3$
to a four electron
one ($7s5f^3$, $7p5f^3$ or $6d5f^3$)
and opens a possibility to obtain a nonmagnetic ground state without
invoking the concept of the octupole order parameter.\cite{San2}
From this point of view it represents an alternative to the latter,
and we believe that both approaches deserves a thorough
experimental consideration and verification.

Our finding that the CEF splitting in NpO$_2$ is rather small 
($\sim 50-150$~K) contrasts the commonly accepted value of 55~meV. \cite{Amo} 
We think that the excitation at 55~meV
observed experimentally by neutron scattering \cite{Amo} is a higher
excitation of the neptunium electronic complex possibly
involving the valence electrons on the Np-O bonds.
The scale of CEF excitations calculated in the present work
is a better match for the width of peaks in magnetic  neutron scattering
cross sections. \cite{Amo} 
We believe that the crystal and mean field
should be of the same order of magnitude and the transition temperature
of 25~K gives a natural estimation for it.
An implicit support of our viewpoint is the
fact that in the ordered phase there appears an inelastic peak
centered at about 6.4 meV. \cite{Amo}

The phase transition to the ordered phase sets in at 25 K
and reduces the basic local symmetry of the neptunium sites to
$S_6=C_3 \times i$ (or ${\bar 3}$). This symmetry
holds for both $Pn{\bar 3}m$ and $Pa{\bar 3}$ space groups.
[The addition of three two-fold axes or three 
mirror planes for $Pn{\bar 3}m$ changes
the point group to $D_{3d}$ (or ${\bar 3}2$) but this is not
of principal importance.]
We have found that the structural change can not be ascribed
to the bilinear quadrupole-quadrupole Coulomb repulsion,
which is too weak to drive the transition. Therefore, 
one has to resort to the superexchange
interactions \cite{super} via oxygen as a driving force for 
the transformation.
In this work we have introduced an effective (enhanced) quadrupole 
interaction and studied the interplay between it and the crystal electric
field.
The influence of both interactions on the transition temperature
has been investigated in detail for the $5f^3$ configuration, Sec.~IV 
and figures 3, 4, and 5.

The most intriguing question is the disappearance of the neptunium
magnetic moments below 25~K.
This however can be explained if we consider a four electron complex
at the neptunium site. We have demonstrated that all relevant
configurations ($7s5f^3$, $7p5f^3$, $6d5f^3$) can lead to 
a nonmagnetic ground state separated
from the magnetic excitations by an energy gap 
larger than $\sim 25$~K, Sec.~V. Perhaps, the most clear example
is the $7p5f^3$ configuration (Tables \ref{table12'}, \ref{table14}). 
In the disordered phase
the ground state is a triplet (Table \ref{table12'}), while in 
the ordered phase
it becomes a singlet (Table \ref{table14}).
The general idea for the loss of magnetic moments is
similar to the one used by Kondo and Anderson
and often referred to as the Kondo effect. \cite{Ful}
Notice however, that here we are dealing with the intrasite 
interactions treated
on {\it ab initio} level.
In particular, we replace the Anderson hybridization \cite{And2}
which is {\it linear} in terms of 
creation/annihilation operators for valence and localized
electrons, by the Coulomb intrasite repulsion, which being
a density-density coupling is {\it bilinear}
in terms of these operators.
Another important theoretical ingredient 
of our model is the symmetry lowering which modifies the
excitation spectrum of the electron system at low temperatures. 
This part
is absent in the Kondo mechanism.

CEF and mean field have been objects of theoretical interest
for many years \cite{Ste,Hut,New1,New2}
and we would like to mention here some 
important relations between
our model and other approaches. We have shown that
CEF effects can be perceived as a first meaningful
term of the intersite multipole expansion, when all neighbors of 
a neptunium site
are considered in the spherical approximation ($l'=0$).
It is then reduced to a single particle potential. \cite{New1,New2}
The intersite nonspherical terms are also included in the 
full potential (FP) electron band structure calculations
like FP-LMTO (linear muffin-tin orbital method)
and FP-LAPW (linear augmented plane wave method). \cite{Sin}
Therefore, in principle
one can say that the CEF effects are equivalent
to the full potential treatment. \cite{Wei,ND}
However, there are two very important caveats here.
First, in the band structure calculations the nonspherical
terms of the potential apply
to itinerant electrons in the ground state, 
while CEF effects are considered
usually for localized electrons in the ground and {\it excited} 
states.
The second more important objection is that practically all 
band structure calculations
are based on the single determinant approximation.
This intrinsic feature does not allow to describe the 
intrasite interactions fully. 
In particular, the atomic term structure and Hund's rules 
are excluded from the
consideration. This shortcoming does not apply to our
treatment which is based on a many determinant approach.
For the intrasite part of interactions our model is very close
to the scheme described by Condon and Shortley for the
electron spectra of atoms and ions although there are some
unimportant differences. 
The main approximation and limitation of our approach is
the choice of basis. Once the many electron basis states are fixed,
the work of calculating matrix elements is reduced mainly
to classifications of all possible permutations, which is done
numerically. 
Thus, the fundamental group of electron permutations \cite{Ell}
is explicitly taken into account. 
No additional approximations used sometimes 
for crystal field calculations are made.
This distinguishes our approach from Stevens' \cite{Ste}
where the CEF is expressed
in terms of equivalent
operators $J_x$, $J_y$, and $J_z$. The latter approach as well
as the work of Lea, Leask and Wolf for cubic CEF \cite{Lea}
based on it are justified only if $J$ is a good quantum number. 
Notice also that the approach of Stephens starts with the symmetry
arguments while the interactions are introduced later in 
a phenomenological manner.

However, the present calculation scheme does not take
into account chemical bonding in an
intrinsic way. Therefore, further development of the
method should be focused on this problem.

\acknowledgments 

We thank C. Detlefs, P. Santini and other authors of 
Ref.~\onlinecite{Pai} for helpful discussions and
communications on this problem.
This work has been financially supported by
the Fonds voor Wetenschappelijk Onderzoek, Vlaanderen.

\appendix

\section{Bilinear quadrupole interactions for $Pn{\bar 3}m$ 
and $Pa{\bar 3}$}

Here we calculate the direct bilinear electronic 
quadrupole-quadrupole interactions for $Pn{\bar 3}m$ 
and $Pa{\bar 3}$ structures, Fig.~\ref{fig_struc}.
We consider the quadrupolar components $S_i$
of $T_{2g}$ symmetry ($i=1-3$) at site 
$\vec{n}=0 \equiv (0,0,0)$ ($\vec{n} \in \{ n_1 \}$).
There are 12 nearest neighbors of $\vec{n}$ belonging
to three different sublattices $\{ n_p \}$, $=2-4$.
The interactions between three components $S_i$ 
centered at $\vec{n}=0$ and those ($S_j$) located at four nearest
neighbors ($\vec{n}'=1-4$) of the second sublattice 
($\vec{n}' \in \{ n_2 \}$) are given in Table \ref{table_app}.
%
\begin{table} 
\caption{ 
The matrix of interaction $S_i(\vec{n})-S_j(\vec{n}')$ 
between three quadrupolar components
of $T_{2g}$ symmetry, $S_1=Y_2^{1s}$, $S_2=Y_2^{1c}$, $S_3=Y_2^{2s}$.
$S_i$ are centered at 
$\vec{n}=(0,0,0)$, $S_j$ at four neighbors $\vec{n}'$ of 
the second sublattice $\{ n_2 \}$.
\label{table_app}     } 
\begin{ruledtabular}
 \begin{tabular}{c l c c c c c c r} 
$\vec{n}'$ & Coord. & $(i,j)=$ & $(1,1)$ & (2,2) & (3,3) & (1,2) & (1,3) 
          & (2,3) \\
\hline 
 1 & $a(0,\frac{1}{2},\frac{1}{2})$ & & $\gamma$ & $\alpha$ & $\alpha$ & 0 & 0
                                    & $\beta$ \\
 2 & $a(0,-\frac{1}{2},\frac{1}{2})$ & & $\gamma$ & $\alpha$ & $\alpha$ & 0 & 0
                                     & $-\beta$ \\
 3 & $a(0,-\frac{1}{2},-\frac{1}{2})$ & & $\gamma$ & $\alpha$ & $\alpha$ & 0 & 0
                                      & $\beta$ \\
 4 & $a(0,\frac{1}{2},-\frac{1}{2})$ & & $\gamma$ & $\alpha$ & $\alpha$ & 0 & 0
                                     & $-\beta$ 
\end{tabular} 
\end{ruledtabular}
\end{table} 

The matrix $S'_i(\vec{n}) - S'_j(\vec{n}')$ 
for the fourth sublattice $\{ n_4 \}$ 
is given by the same Table provided that the cyclic permutation
$S_1 \rightarrow S'_3$, $S_2 \rightarrow S'_1$,
$S_3 \rightarrow S'_2$ is performed.
Here we label four nearest neighbors $\vec{n}'=5-8$ of $\vec{n}$ as 
 $5 \equiv a(1/2,1/2,0)$, $6 \equiv a(-1/2,1/2,0)$, $7 \equiv a(-1/2,-1/2,0)$,
 and $8 \equiv a(1/2,-1/2,0)$.
The matrix $S''_i(\vec{n}) - S''_j(\vec{n}')$
for the third sublattice $\{ n_3 \}$ 
is obtained from Table \ref{table_app} by replacing 
$S_1 \rightarrow S''_2$, $S_2 \rightarrow S''_3$,
$S_3 \rightarrow S''_1$. Here $\vec{n}'$ runs over
the sites $9 \equiv a(1/2,0,1/2)$, $10 \equiv a(-1/2,0,1/2)$, 
$11 \equiv a(-1/2,0,-1/2)$, and $12 \equiv a(1/2,0,-1/2)$.

For the $Pn{\bar 3}m$ order of quadrupoles we have
${\cal S}_a$, Eq.~(\ref{3c.1a}), at $(0,0,0)$ and 
${\cal S}_d$, Eq.~(\ref{3c.1d}), for $\vec{n}'=1-4$.
By means of Table \ref{table_app} we find that this interaction
is given by
\begin{eqnarray}
  E^{QQ}_0(\{n_2\}) = \frac{4}{3}(\gamma-2\alpha) .
\label{a.1} 
\end{eqnarray}
The same expression is obtained for the interactions
with four neighbors of third and forth
sublattices, $E^{QQ}_0(\{n_3\})$ and $E^{QQ}_0(\{n_4\})$. 
The total interaction then is
\begin{eqnarray}
  E^{QQ}(Pn{\bar 3}m) &=& E^{QQ}_0(\{n_2\})+
  E^{QQ}_0(\{n_3\})+E^{QQ}_0(\{n_4\}) \nonumber \\ 
  &=& 4(\gamma-2\alpha) > 0 ,
\label{a.2} 
\end{eqnarray}
because $\gamma > 0$ and $\alpha < 0$
independently of the lattice constant $a$.

For the $Pa{\bar 3}$ structure we have the function ${\cal S}_b$,
Eq.~(\ref{3c.1b}), at $\vec{n}'=1-4$, and 
$E^{QQ}_0(\{n_2\}) = -4\gamma /3$. The same value is obtained
for $E^{QQ}_0(\{n_3\})$ and $E^{QQ}_0(\{n_4\})$.
As a result we arrive at
\begin{eqnarray}
  E^{QQ}(Pa{\bar 3}) = -4 \gamma < 0 .
\label{a.3} 
\end{eqnarray}

We conclude that the $Pn{\bar 3}m$ structure always leads to a repulsion
while the $Pa{\bar 3}$ to an attraction.

\section{Correction of Slater integrals for Np}

Since the experimental data on the energy
splittings of the $4f^3$ configuration of 
Pr$^{3+}$ and Nd$^{4+}$
are available from the atomic database of NIST, Ref.\ \onlinecite{La_sp},
while there is no such information for $5f^3$, we have
performed calculations of $v^{F-F}_l$ ($l=2,4,6$), Eq.~(\ref{2.18a}), 
by using the radial dependence of ${\cal R}_f$ obtained from 
LDA calculations of atoms. After this we diagonalize the
Hamiltonian of the free ion ($V^{(3)}+H_{so}$)
and compared our calculated spectra with the experimental ones.
We have noticed that the comparison is improved 
(the sequence of terms corresponds to the experimental one) 
if we reduce 
$v^{F-F}_2$ and $v^{F-F}_6$ by a factor of 0.75 while keep $v^{F-F}_4$
almost the same (factor 0.975).
Therefore, we have used the same scale factors for Np in
NpO$_2$ and obtained parameters given by Eq.~(\ref{3.2}).



\end{document}